%% file: TOP-19-006_temp.tex
\pdfoutput=1
\documentclass[11pt,twoside,a4paper,cmspaper,final,collab]{cms-tdr}
\def\svnVersion{19e0bd8}\def\svnDate{2022/05/02}\def\cmsCernNoTag{CERN-EP-2021-248}\def\cmsCernDate{\today}\def\cmsMessage{Published in the Journal of High Energy Physics as \href{http://dx.doi.org/10.1007/JHEP06(2022)082}{\doi{10.1007/JHEP06(2022)082}.}}
\begin{document}\cmsNoteHeader{TOP-19-006}

\newlength\cmsTabSkip\setlength{\cmsTabSkip}{1ex}
\newcommand{\ttzw}{\ensuremath{\ttbar\text{+\PZ/\PW}}\xspace}
\newcommand{\xsectheo}{\ensuremath{832^{+20}_{-29}{\,\text{(scale)}} \pm 35\,({\mathrm{PDF}}+\alpS)\unit{pb}}\xspace}
\newcommand{\xsectheoProductionVecU}{\ensuremath{634^{+113}_{-90} \pm 8}\xspace}
\newcommand{\xsectheoProductionVecC}{\ensuremath{58^{+9}_{-7} \pm 8}\xspace}
\newcommand{\xsectheoDecayVec}{\ensuremath{32.0^{+0.8}_{-1.1} \pm 1.3}\xspace}
\newcommand{\xsectheoProductionScaC}{\ensuremath{12.1^{+2.0}_{-1.6} \pm 1.8}\xspace}
\newcommand{\xsectheoProductionScaU}{\ensuremath{139^{+26}_{-20} \pm 2}\xspace}
\newcommand{\xsectheoDecaySca}{\ensuremath{4.0^{+0.1}_{-0.1} \pm 0.2}\xspace}
\newcommand{\xsectheoProductionTenU}{\ensuremath{2908^{+503}_{-401} \pm 37}\xspace}
\newcommand{\xsectheoProductionTenC}{\ensuremath{292^{+42}_{-35} \pm 37}\xspace}
\newcommand{\xsectheoDecayTen}{\ensuremath{187^{+5}_{-6} \pm 8}\xspace}
\newcommand{\mt}{\ensuremath{m_{\PQt}}\xspace}
\newcommand{\pp}{\ensuremath{\Pp\Pp}\xspace}
\newcommand{\Wjets}{\ensuremath{\PW\text{+jets}}\xspace}
\newcommand{\tw}{\ensuremath{\PQt\PW}\xspace}
\newcommand{\cvector}{\ensuremath{\text{C}_{\text{vector}}\xspace}}
\newcommand{\cscalar}{\ensuremath{\text{C}_{\text{scalar}}\xspace}}
\newcommand{\ctensor}{\ensuremath{\text{C}_{\text{tensor}}\xspace}}
\newcommand{\ovector}{\ensuremath{O_{\text{vector}}\xspace}}
\newcommand{\oscalar}{\ensuremath{O_{\text{scalar}}\xspace}}
\newcommand{\otensor}{\ensuremath{O_{\text{tensor}}\xspace}}
\newcommand{\olqthree}{\ensuremath{O_{\text{lq}}^{(3)abcd}\xspace}}
\newcommand{\olqone}{\ensuremath{O_{\text{lq}}^{(1)abcd}\xspace}}
\newcommand{\olu}{\ensuremath{O_{\text{lu}}^{abcd}\xspace}}
\newcommand{\oeq}{\ensuremath{O_{\text{eq}}^{abcd}\xspace}}
\newcommand{\oeu}{\ensuremath{O_{\text{eu}}^{abcd}\xspace}}
\newcommand{\olequone}{\ensuremath{O_{\text{lequ}}^{(1)abcd}\xspace}}
\newcommand{\olequthree}{\ensuremath{O_{\text{lequ}}^{(3)abcd}\xspace}}
\newcommand{\antikt}{anti-\kt}
\newcommand{\cxlbd}{\text{C}_x/\Lambda^2 = 1\TeV^{-2}}
\providecommand{\cmsTable}[1]{\resizebox{\textwidth}{!}{#1}}

\cmsNoteHeader{TOP-19-006}
\title{Search for charged-lepton flavor violation in top quark production and decay in \texorpdfstring{\pp collisions at $\sqrt{s} = 13\TeV$}{pp collisions at sqrt(s)=13 TeV} }

\date{\today}
\abstract{
Results are presented from a search for charged-lepton flavor violating (CLFV) interactions in top quark production and decay in \pp collisions at a center-of-mass energy of 13\TeV. The events are required to contain one oppositely charged electron-muon pair in the final state, along with at least one jet identified as originating from a bottom quark. The data correspond to an integrated luminosity of 138\fbinv, collected by the CMS experiment at the LHC. This analysis includes both the production ($\PQq \to \Pe \PGm \PQt$) and decay ($\PQt \to \Pe \PGm \PQq$) modes of the top quark through CLFV interactions, with \PQq referring to a \PQu or \PQc quark. These interactions are parametrized using an effective field theory approach. With no significant excess over the standard model expectation, the results are interpreted in terms of vector-, scalar-, and tensor-like CLFV four-fermion effective interactions. Finally, observed exclusion limits are set at 95\% confidence levels on the respective branching fractions of a top quark to an $\Pe\PGm$ pair and an up (charm) quark of  $0.13\times 10^{-6}$ ($1.31\times 10^{-6}$), $0.07\times 10^{-6}$ ($0.89\times 10^{-6}$), and $0.25\times 10^{-6}$ ($2.59\times 10^{-6}$) for vector, scalar, and tensor CLFV interactions, respectively.
}

\hypersetup{
pdfauthor={CMS Collaboration},
pdftitle={Search for charged-lepton flavor violation in top quark production and decay in pp collisions at sqrt(s) = 13 TeV},%
pdfsubject={CMS},
pdfkeywords={CMS, top quark}}
\maketitle

\section{Introduction}
\label{sec:introduction}

The framework of the standard model (SM) with massless neutrinos contains intrinsic global symmetries such as those involving individual lepton flavor quantum numbers.
As a consequence, the mixing of neutrino flavors is forbidden and the flavor of charged leptons cannot be changed in weak interactions. 
The discovery of neutrino oscillations proved that neutrinos are massive particles and that lepton flavor is not always conserved in the neutral-lepton sector \cite{Choudhury:2018byy}. 
Neutrino oscillations also give rise to charged-lepton flavor violating (CLFV) processes; however, these processes are highly suppressed because of the small values of neutrino masses that are far below experimental sensitivity.
Any evidence for such rare processes would therefore serve as a clear signature of physics beyond the SM.

There are many theoretical scenarios extending the SM, such as the two-Higgs doublet model \cite{DiazCruz:1999xe}, the minimal supersymmetric model \cite{Crivellin:2018mqz}, and the inverse seesaw model \cite{Malinsky:2009df}, 
under which the CLFV rate can be close to the current experimental sensitivity, and therefore be accessible to study.
Searches for CLFV processes can be divided into low- and high-energy categories \cite{Calibbi:2017uvl}.
The most promising low-energy channels are $\PGm \to \Pe\PGg$, $\PGm \to \Pe\Pe\Pe$, and $\PGm \to \Pe$ conversion in nuclei, as well as similar CLFV processes involving $\PGt$ leptons \cite{Adam:2013mnn,Mihara:2013zna}.
The CERN LHC provides the highest sensitivity to high-energy CLFV processes involving a heavy particle, such as the \PZ boson, Higgs boson, or the top quark.
In this context, the ATLAS and CMS experiments have performed searches for CLFV decays of the \PZ boson in $\Pe\PGm$, $\Pe\PGt$, and $\PGm\PGt$ final states and of the Higgs boson in $\Pe\PGm$, $\Pe\PGt$, and $\PGm\PGt$ channels in \pp collisions at 13\TeV, finding no significant excess of events over the expected SM background \cite{Aaboud:2018cxn,Aad:2016blu,Sirunyan:2017xzt, CMS:2021rsq, CMS:2016cvq, ATLAS:2019old}.

In the past few years, different measurements of B meson decays that involve leptons have hinted at the presence of possible small violations of lepton universality \cite{Belle:2015pkj,LHCb:2021trn}. 
It has been pointed out \cite{Glashow:2014iga} that models accommodating such levels of violation of lepton universality  generally also lead to observable effects in lepton flavor violation.
Moreover, physics models with solutions to the possible anomalies seen in the bottom quark sector predict similar effects in the top quark sector \cite{Bissmann:2020mfi}.
For example, certain leptoquark models can accommodate the observed deviation in the measurements of the branching fraction ratios of $\overline{\PB}\to {\PD}^*\PGtm\overline{\PGn}_{\PGt}$ relative to $\PAB\to {\PD}^*\ell^{-}\PSGn_{\PGt}$ (where $\ell = \Pe$ or $\PGm$) \cite{Dumont:2016xpj,Kim:2018oih}. These models would imply branching fractions of $\PQt\to \ell \ell^\prime \PQc$ reaching $\approx 10^{-6}$, with $\ell$ and $\ell^\prime$ representing different-flavor charged leptons.
Searching for CLFV processes related to the top quark could therefore  shed light on anomalies seen in \PB meson decays.

Assuming the mass scale of new physics responsible for CLFV processes is larger than the energy scale directly accessible at the LHC, CLFV interactions of top quarks are described through an effective Lagrangian consisting of dimension-six operators ($O_x$) weighted by the Wilson coefficients ($\text{C}_x$) over powers of the new mass scale ($\Lambda$),

\begin{equation}
  \mathcal{L} = \mathcal{L}_\text{SM} + \mathcal{L}_\text{eff} = 
    \mathcal{L}_\text{SM} + \sum_x \frac{\text{C}_x}{\Lambda^2} O_x + \cdots
\end{equation}

In the Warsaw basis of dimension-six operators, the following operators give rise to top quark CLFV interactions \cite{AguilarSaavedra:2018nen}:
\begin{align}
\label{eq0:1}
    \olqthree&
    = (\PAl_a\PGg^\mu \tau^I \Pl_b)
      (\PAQq_c\PGg_\mu \tau^I \PQq_d)
    ,\\
\label{eq0:2}
    \olqone&
    =(\PAl_a\PGg^\mu \Pl_b)
      (\PAQq_c\PGg_\mu \PQq_d)
    ,\\
\label{eq0:3}
    \olu~~~&
    =(\PAl_a\PGg^\mu \Pl_b)
      (\PAQu_c\PGg_\mu \PQu_d)
    ,\\
\label{eq0:4}
    \oeq~~~&
    =({\overline{\Pe}}_a\PGg^\mu \Pe_b)
      (\PAQq_c\PGg_\mu \PQq_d)
    ,\\
\label{eq0:5}
    \oeu~~~&
    =({\overline{\Pe}}_a\PGg^\mu \Pe_b)
      (\PAQu_c\PGg_\mu \PQu_d)
    ,\\
\label{eq0:6}
    \olequone&
    =(\PAl_a \Pe_b)\;\varepsilon\;
      (\PAQq_c \PQu_d)
    ,\\
\label{eq0:7}
    \olequthree&
    =(\PAl_a \sigma^{\mu\nu} \Pe_b)\;\varepsilon\;
      (\PAQq_c \sigma_{\mu\nu} \PQu_d),
\end{align}
where  $a\neq b$ are lepton-flavor indices, $c$ and $d$ are quark-flavor indices, \PQq and \Pl represent left-handed fermion doublets, \PQu and \Pe the right-handed fermion singlets, $\tau^I$ the Pauli matrices, $\varepsilon\equiv i\tau^2$ is the antisymmetric $SU(2)$ tensor, $\sigma^{\mu\nu}= \frac{i}{2} [\gamma^{\mu},\gamma^{\nu}]$, and $\gamma^{\mu}$ the Dirac matrices.
To reduce the number of free parameters, we assume that all Wilson coefficients obtained from the permutation of the lepton and quark flavors for a fixed set of $abcd$ are equal.  
It can be shown that the part of the $\olqthree$  operator that contributes to top quark CLFV processes has Lorentz structure analogous to the $\olqone$ operator \cite{Durieux:2014xla}. 
The $\olqthree$ operator is therefore not included in this analysis. 
The operators in Eqs. (\ref{eq0:2})--(\ref{eq0:7}) are classified on the basis of their Lorentz structure as vector ($\olqone$, $\olu$, $\oeq$, and $\oeu$), scalar ($\olequone$), and tensor ($\olequthree$) operators. 
These CLFV vector, scalar, and tensor operators, denoted by $\ovector$, $\oscalar$, and $\otensor$, respectively, are given by:
\begin{align}
\label{eq2:1}
    \ovector
    &=
    O_{\Pl\PQq}+
    O_{\Pl\PQu}+
    O_{\Pe\PQq}+
    O_{\Pe\PQu}
    ,\\
\label{eq2:2}
    \oscalar
    &=
    O_{\Pl\Pe\PQq\PQu}^{(1)} + \text{h.c}
    ,\\
\label{eq2:3}
    \otensor
    &=
    O_{\Pl\Pe\PQq\PQu}^{(3)} + \text{h.c},
\end{align}
where $\ovector$ represents the sum of the operators in Eqs. (\ref{eq0:2})--(\ref{eq0:5}).

\begin{figure}[htb!]
  \begin{center}
      \includegraphics[width=0.95\textwidth]{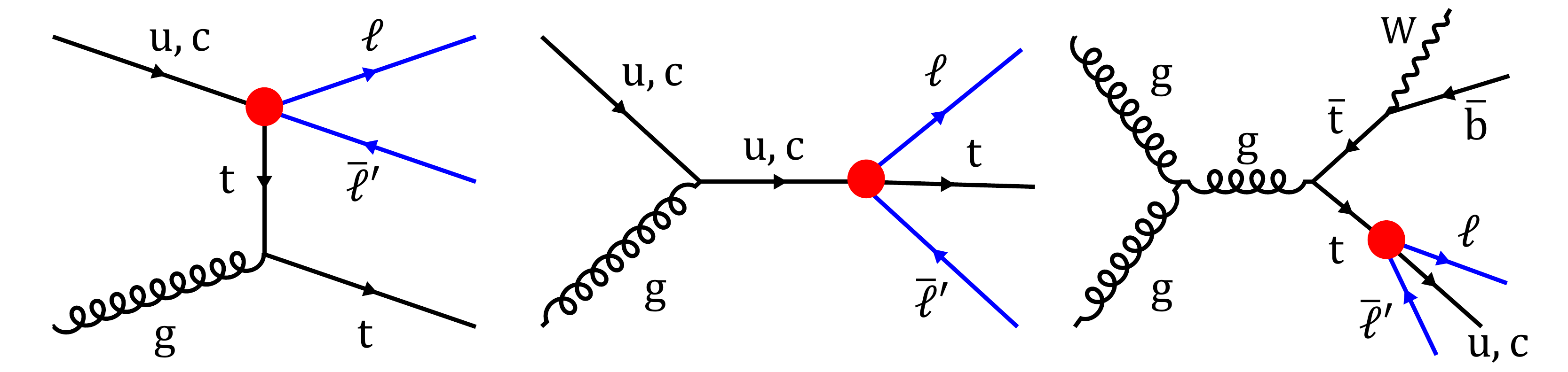}
    \caption{ Feynman diagrams for single top quark production (left and middle) and top quark decays in SM \ttbar events (right) via CLFV interactions. The CLFV vertex is marked as a filled circle. 
    \label{Feyn}}
  \end{center}
\end{figure}

We probe three Wilson coefficients related to the operators, $\cvector$, $\cscalar$, and $\ctensor$.
The operators in Eqs. (\ref{eq2:1})--(\ref{eq2:3}) can lead to four-fermion interactions involving the top quark, the up or charm quark, and two leptons of different flavor.
These four-fermion interactions open new top quark decay modes, \eg, $\PQt\to \ell \ell^\prime \PQq$,  where $\ell$ and $\ell^\prime$ are charged leptons with different flavors, and q is a $\PQu$ or $\PQc$ quark \cite{Davidson:2015zza}.
In addition to top quark decays, CLFV interactions at the LHC contribute to single top quark production in association with a pair of leptons of different flavor.
Figure~\ref{Feyn} displays representative Feynman diagrams for single top quark production and decay of the top quark in top quark-antiquark pair production (\ttbar) via CLFV interactions.

Final-state signatures are determined by the lepton flavors and decay modes of the \PW boson from top quark decays. The \PW boson can decay either leptonically to a charged lepton and a neutrino or to two quarks that develop into jets via quantum chromodynamics (QCD) processes. 
Final states in which \PW bosons decay into quarks have cross sections larger than for leptonic decays.  
This analysis combines first searches for ``$\Pe\PGm \PQt\PQu$'' and  ``$\Pe\PGm \PQt\PQc$'' CLFV interactions in top quark production with decays to the $\Pe\PGm$ final state at $\sqrt{s} = 13\TeV$. 
We select signal events containing an  oppositely charged $\Pe\PGm$ pair and a top quark that decays fully hadronically.
The data used in the analysis correspond to an integrated luminosity of 138\fbinv, collected by the CMS experiment at the LHC during 2016--2018.
The top quark production mode via CLFV interactions plays a leading role in the sensitivity of the search compared to the decay mode. 
The result is interpreted in terms of limits on vector, scalar, and tensor four-fermion interactions originating from dimension-six operators within the framework of effective field theory.
 
The paper is organized as follows. Section~\ref{sec:cmsdetector} describes the main features of the CMS detector.  
Section~\ref{sec:samples} provides the details of the Monte Carlo (MC) simulations of signal and background.
The event reconstruction is outlined in Section~\ref{sec:Eventselection}.  
In Section~\ref{sec:SignalExtraction}, we discuss the distinctive features of signal relative to background, followed by a description of the signal extraction.  
Systematic uncertainties are discussed in Section~\ref{sec:unc}.  
Section~\ref{sec:results} presents the results, and Section~\ref{sec:summary} provides a summary of the paper.

\section{The CMS detector}
\label{sec:cmsdetector}

The central feature of the CMS apparatus is a superconducting solenoid of 6\unit{m} internal diameter, providing a magnetic field of 3.8\unit{T}. 
Within the solenoid volume are a silicon pixel and strip tracker, a lead tungstate crystal electromagnetic calorimeter (ECAL), and a brass and scintillator hadron calorimeter, each composed of a barrel and two endcap sections.
Forward calorimeters extend the pseudorapidity coverage provided by the barrel and endcap detectors. Muons are measured in gas-ionization detectors embedded in the steel flux-return yoke outside the solenoid. 
A more detailed description of the CMS detector, together with a definition of the coordinate system used and the relevant kinematic variables, can be found in Ref.~\cite{Chatrchyan:2008zzk}. 

Events of interest are selected using a two-tiered trigger system. The first level, composed of custom hardware processors, uses information from the calorimeters and muon detectors to select events at a rate of around 100\unit{kHz} within a fixed latency of about 4\mus~\cite{Sirunyan:2020zal}. The second level, known as the high-level trigger, consists of a farm of processors running a version of the full event reconstruction software optimized for fast processing, and reduces the event rate to around 1\unit{kHz} before data storage~\cite{Khachatryan:2016bia}.

\section{Simulation of background and signal}
\label{sec:samples}

Monte Carlo events are used to estimate the SM backgrounds and samples are simulated through independent events generated for the years 2016, 2017, and 2018 so as to match  the different data-taking conditions.
The SM \ttbar, single top quark production in association with a \PW boson ($\tw$), and diboson events (including $\PW\PW$, $\PW\PZ$, and $\PZ\PZ$) are  simulated at next-to-leading order (NLO) using the \POWHEG v2 event generator~\cite{bib:powheg0,bib:powheg,bib:powheg2, Frixione:2007nw}.
All other background processes, including Drell--Yan processes produced with additional jets, a \PW boson with additional jets (\Wjets), and \PW or \PZ bosons produced in association with \ttbar (\ttzw) are simulated using the \MGvATNLO v2.4.2 (v2.2.2 for 2016) generator~\cite{Alwall:2014hca}.

The cross sections  are  calculated  at  the  highest  orders  of  perturbative  QCD  currently  available.
This corresponds to next-to-NLO (NNLO) for Drell--Yan and \Wjets~\cite{Li:2012wna}, approximate NNLO for single top quark in the \tw channel~\cite{bib:twchan}, and NLO calculations for diboson~\cite{bib:mcfm:diboson} and \ttzw~\cite{bib:Maltoni:2015ena}. 
The SM \ttbar events are  normalized to their NNLO cross sections (\xsectheo) calculated with the \textsc{Top++}2.0 program~\cite{Czakon:2011xx}, where PDF is the parton distribution function and \alpS is the strong coupling constant, assuming a top quark mass of 172.5\GeV. 
To improve the modeling of the transverse momentum (\pt) spectrum of the top quark in \POWHEG, simulated SM \ttbar events are weighted as a function of the \pt of the top quark to match the expectations at NNLO QCD accuracy, including electroweak corrections~ \cite{Czakon:2017wor}.

The effective Lagrangian extracted from the operators defined in Eqs. (\ref{eq0:2})--(\ref{eq0:7}) is implemented in the \textsc{FeynRules} program \cite{Brivio:2017btx,Dedes:2019uzs}, and then used in the \MGvATNLO generator for the cross section calculation and event generation at leading order. 
The top quark CLFV signal has two components: (i) events from  the production of SM \ttbar followed by a CLFV decay of one of the top quarks, and (ii) single top quark production in association with an $\Pe\PGm$ pair via CLFV interactions, as shown in Fig.~\ref{Feyn}.
Due to the fact that single top quark production via $\Pe\PGm\PQt\PQu$ CLFV interactions is initiated by a u quark,
and u quarks are mostly proton valence quarks with a very different Bjorken-$x$ spectrum relative to sea quarks,
the production rate and kinematic distributions of final-state particles are different than when a sea quark is involved in the interaction, as is the case for single top quark production via $\Pe\PGm\PQt\PQc$ CLFV.
Each component of signal is therefore generated independently for the $\Pe\PGm\PQt\PQu$ and $\Pe\PGm\PQt\PQc$ CLFV interactions.
Events from the ``$\Pe\PGt \PQt\PQq$'' and  ``$\PGm\PGt \PQt\PQq$'' CLFV interactions are not included in the signal samples.
Since there is no interference between the SM and the signal processes, signal events are generated separately from the SM background.
The new mass scale and the Wilson coefficients are arbitrarily chosen to be $\Lambda=1\TeV$ and $\text{C}^{\Pe\PGm \PQt \PQq}_{x}=1$  for event generation.
For SM \ttbar production with top quark CLFV decay, the cross section is calculated using the SM \ttbar cross section at NNLO times the branching fraction $\mathcal{B}(\PQt\to \Pe \PGm \PQq)$, assuming $\Lambda=1$ \TeV and $\text{C}^{\Pe\PGm \PQt \PQq}_{x}=1$ for both the u and c quarks~\cite{Kile:2008rp}.
Theoretical cross sections, for single top quark
production and top quark decays via the vector, scalar, and tensor CLFV interactions are shown in table \ref{tab:CrossSec}.

\begin{table}[!ht]
\topcaption{Theoretical cross sections, in fb, for single top quark
production and top quark decays via the vector, scalar, and tensor CLFV interactions, assuming a top quark mass of 172.5\GeV, the top quark decay width 1.33\GeV, $\Lambda=1\TeV$ and $\text{C}^{\Pe\PGm \cPqt \cPq}_{x}=1$. The uncertainties from the QCD scales and PDF are given ($\sigma^{+\text{scale}}_{-\text{scale}}\pm$PDF).
\label{tab:CrossSec}}
\centering
\begin{tabular}{lccc}
channel & Vector & Scalar & Tensor \\
\hline
Production ($\Pe\PGm \PQt\PQu$) & \xsectheoProductionVecU    & \xsectheoProductionScaU & \xsectheoProductionTenU    \\
Production ($\Pe\PGm \PQt\PQc$)  & \xsectheoProductionVecC    & \xsectheoProductionScaC & \xsectheoProductionTenC    \\
Decay ($\Pe\PGm \PQt \PQq$) & \xsectheoDecayVec    & \xsectheoDecaySca & \xsectheoDecayTen    \\
\end{tabular}
\end{table}

The NLO PDF sets, NNPDF3.0 \cite{Ball:2014uwa}, are used in the generation of MC events collected in 2016, while the NNLO PDF sets from NNPDF3.1 \cite{Ball:2017nwa} are used for the 2017--2018 data.
Parton showering and hadronization are handled through \PYTHIA v8.205~\cite{Sjostrand:2014zea} using the underlying-event tune CUETP8M1~\cite{Khachatryan:2015pea}  for 2016 data and tune CP5~\cite{Sirunyan:2019dfx} for 2017-2018.
For SM \ttbar production, the CP5 tune is also used for 2016 data. Simulated minimum-bias events are added to the MC simulations to model the impact of additional \pp interactions within the same or adjacent bunch crossing (pileup).
Simulated events are then reweighted  to reproduce the pileup distribution observed in data.  
All generated events undergo a full simulation of the detector response using \GEANTfour~\cite{Agostinelli:2002hh}.

\section{Event selection}
\label{sec:Eventselection}

Signal events contain an oppositely charged  $\Pe\PGm$ pair together with multiple jets,
one of which is expected to stem from the hadronization of a bottom quark that originates from the $\PQt \to \PQb\PW$ decay.
The data for this analysis are collected using a combination of triggers designed to record events containing a single muon, a single electron, or an $\Pe\PGm$ pair passing isolation and identification criteria.
For the single-electron (muon) trigger, at least one electron (muon) with \pt larger than 27, 35, and 32 (24, 27, and 24) \GeV is required for 2016, 2017, and 2018 data, respectively.
The $\Pe\PGm$ trigger selects events having an electron with $\pt > 12\GeV$ and a muon with $\pt > 23\GeV$, or an electron with $\pt > 23\GeV$ and a muon with $\pt > 8\GeV$, in all years.
The trigger efficiency within the detector acceptance is measured in data to be greater than 96\%  for events with at least an $\Pe\PGm$ pair.

Events selected at the trigger level are reconstructed offline using the particle-flow (PF) algorithm \cite{Sirunyan:2017ulk}, 
which identifies and reconstructs each individual particle in an event through an optimized combination of information from the various components of the CMS detector.
The candidate vertex with the largest value of summed physics-object $\pt^2$ is taken 
to be the primary \pp interaction vertex. The physics objects are the jets, 
clustered using the jet finding algorithm (\antikt)~\cite{Cacciari:2008gp,Cacciari:2011ma} with 
the tracks assigned to candidate vertices as inputs, and the associated missing 
transverse momentum, taken as the negative vector sum of the \pt of those jets.
Electron candidates are reconstructed from a combination of a track in the tracker and associated energy deposition in the ECAL~\cite{Khachatryan:2015hwa}.
They are required to have $\pt > 20\GeV$ and to lie within $\abs{\eta}<2.4$, except that
candidates in the transition region between barrel and endcap calorimeters ($1.44 < \abs{\eta} < 1.57$) are removed.
A relative isolation requirement $\text{I}_{\text{rel}} < 0.05$ is imposed where $\text{I}_{\text{rel}}$ 
is the scalar-\pt sum of all neutral and charged hadron, and photon candidates within a distance of $\Delta R = \sqrt{\smash[b]{(\Delta\eta)^2+(\Delta\phi)^2}} < 0.3$ from the axis of the electron candidate, divided by the \pt of the electron candidate.
In addition, stringent electron identification requirements are applied to reject misidentified electron candidates and candidates originating from photon conversions in the detector materials \cite{Khachatryan:2015hwa}.
Muon candidates are reconstructed by associating tracks found in the muon system with tracks in the inner tracking systems~\cite{Sirunyan:2018fpa}.
They are required to have $\pt > 20\GeV$  and $\abs{\eta}<2.4$. 
The relative isolation requirement $\text{I}_{\text{rel}}<0.15$ is applied where $\text{I}_{\text{rel}}$ is calculated for all particles within a cone of radius $\Delta R < 0.4$ from the muon trajectory.
A correction to suppress a residual effect of the pileup is included~\cite{Sirunyan:2018fpa}.
Muon candidates must pass identification requirements \cite{Sirunyan:2018fpa}.
In addition, some dedicated  muon identification requirements are applied to reject misidentified muon candidates of large \pt \cite{Sirunyan:2019yvv}.
Electrons and muons are selected if they are compatible with originating from the primary vertex.

The PF candidates are clustered into jets using the \antikt algorithm with a distance parameter $R = 0.4$.
The charged hadron subtraction procedure \cite{Cacciari:2007fd} mitigates event by event the effect of tracks coming from pileup on the transverse energy of the jet. 
Jets are calibrated in simulation and separately in data, accounting for energy depositions from pileup and from imprecise detector response~\cite{Khachatryan:2016kdb}. 
Jets with $\pt > 30\GeV$ and $\abs{\eta} < 2.4$ are selected for further study.
To prevent overlap between selected jets and selected leptons, jets that are found within a cone of $\Delta R < 0.4$ around any of the selected leptons are removed from the selected set of jets.
Jets originating from the hadronization of bottom quarks are identified (``\PQb tagged'') using deep machine learning algorithms \cite{Sirunyan:2017ezt} with an efficiency of 68\% and a 1\% misidentification rate for gluon and light-flavor quark jets. 
The missing transverse momentum vector \ptvecmiss is computed as the negative vector \pt sum of all the PF candidates in an event, and its magnitude is denoted as \ptmiss~\cite{CMS:2019ctu}.

Events with an oppositely charged  $\Pe\PGm$ pair and at least one jet are selected.
The leading lepton  must have $\pt>25\GeV$.
Events are rejected if the invariant mass of the $\Pe\PGm$ pair is less than 20\GeV~\cite{Sirunyan:2018ucr}. 
Since the top quark CLFV signal has one \PQb quark, events are required to have at least one \PQb tagged jet.

\section{Signal extraction}
\label{sec:SignalExtraction}

The contributions from the SM are estimated using the simulated events introduced in Section~\ref{sec:samples} normalized to the integrated luminosity of the data.
After requiring at least one \PQb tagged jet, the dominant source of background originates from SM \ttbar events, which contribute $\approx$90\% of the total background.
To control this background, events are subdivided according to the number of \PQb tagged jets, irrespective of the number of untagged jets. The signal region includes events with one \PQb tagged jet while events with at least two  such jets are assigned to the \ttbar control region.
The numbers of events with one and greater than one \PQb tagged jets are shown in Table~\ref{cutflow}, together with the expected number of background events in the combined  Run-2  data.
The overall number of events is well described by the expectations in both regions.
In Table~\ref{cutflow}, we give the expected number of events for single top quark  production and top quark decay in the signal channels (cf. Fig.~\ref{Feyn}), assuming $\cxlbd$. 
Signal channels are further categorized by the CLFV interaction (vector, scalar, or tensor) and the $\PQu$ or $\PQc$ quark flavor.

\begin{table}[hp!]
\centering
\topcaption{The number of expected events from SM \ttbar, \tw, and from the other backgrounds; 
and the total background expectation and the number of events observed in data collected during 2016--2018, after all selections in signal (1 \PQb tagged) and control ($>$1 \PQb tagged) regions. 
The total uncertainty, including both statistical and unfited systematic components, is quoted in quadrature for the expected backgrounds.
The expected signal yields for single top quark production and top quark decays via the vector, scalar, and tensor CLFV interactions are also shown with their MC statistical uncertainties, assuming $\cxlbd$. } 
\label{cutflow}
\begin{tabular}{lllllll}
\hline
\multicolumn{3}{l}{Channel} & 1 \PQb tagged & $>$1 \PQb tagged\\
\hline
\multicolumn{3}{l}{\ttbar} & $477 800 \pm 7 900$ & $265 000 \pm 7 100$\\
\multicolumn{3}{l}{\tw} & $49 100 \pm 1 300$ & $7710 \pm 250$\\
\multicolumn{3}{l}{Other} & $7950 \pm 670$ & $850 \pm 70$\\ 
\multicolumn{3}{l}{Total background prediction}  & $534 900 \pm 8 000$ & $273 600 \pm 7 100$\\
\multicolumn{3}{l}{Data}  & $537 236$ & $268 781$\\[2\cmsTabSkip]
\multirow{6}{*}{$\Pe\PGm \PQt\PQu$} & \multirow{2}{*}{Vector} & \PQt decay & $604 \pm 2$ & $45.2 \pm 0.4$\\
&& \PQt production& $17 103 \pm 29$ & $1557 \pm 9$\\
&\multirow{2}{*}{Scalar} & \PQt decay & $78.2 \pm 0.2$ & $6.1 \pm 0.1$\\
&& \PQt production& $3670 \pm 6$ & $336 \pm 2$\\
&\multirow{2}{*}{Tensor} & \PQt decay & $3499 \pm 9$ & $266 \pm 2$\\
&& \PQt production& $61 011 \pm 107$ & $5567 \pm 33$\\[\cmsTabSkip]
\multirow{6}{*}{$\Pe\PGm \PQt\PQc$} & \multirow{2}{*}{Vector} & \PQt decay & $596 \pm 2$ & $90.4 \pm 0.5$\\
&& \PQt production& $1711 \pm 3$ & $166 \pm 1$\\
&\multirow{2}{*}{Scalar} & \PQt decay & $77.7 \pm 0.2$ & $11.4 \pm 0.1$\\
&& \PQt production& $294 \pm 1$ & $28.5 \pm 0.2$\\
&\multirow{2}{*}{Tensor} & \PQt decay & $3467 \pm 8$ & $534 \pm 3$\\
&& \PQt production& $6329 \pm 13$ & $621 \pm 4$\\
\hline
\end{tabular}
\end{table}

The background in the signal  region consists mostly of SM \ttbar events where both \PW bosons decay leptonically.  
Several differences between the signal and the SM \ttbar events are used to construct a discriminating observable.  For example, 
the sources of \ptmiss in signal events are due to detector resolutions, while SM \ttbar events have genuine \ptmiss produced by neutrinos from the \PW boson decays.
Leptons in SM \ttbar events arise from the decay of \PW bosons and have different angular separations and energy spectra relative to signal dilepton events.
Furthermore, signal events have a larger number of light-flavor quark jets because of the multijet top quark decays in signal events. 
To maximize the sensitivity of the search, a boosted decision tree (BDT) that combines several discriminating variables is defined in the  toolkit for multivariate analysis \cite{Hocker:2007ht} and used to distinguish signal from  SM \ttbar events.

\begin{figure}[hp!]
  \begin{center}
      \includegraphics[width=0.44\textwidth]{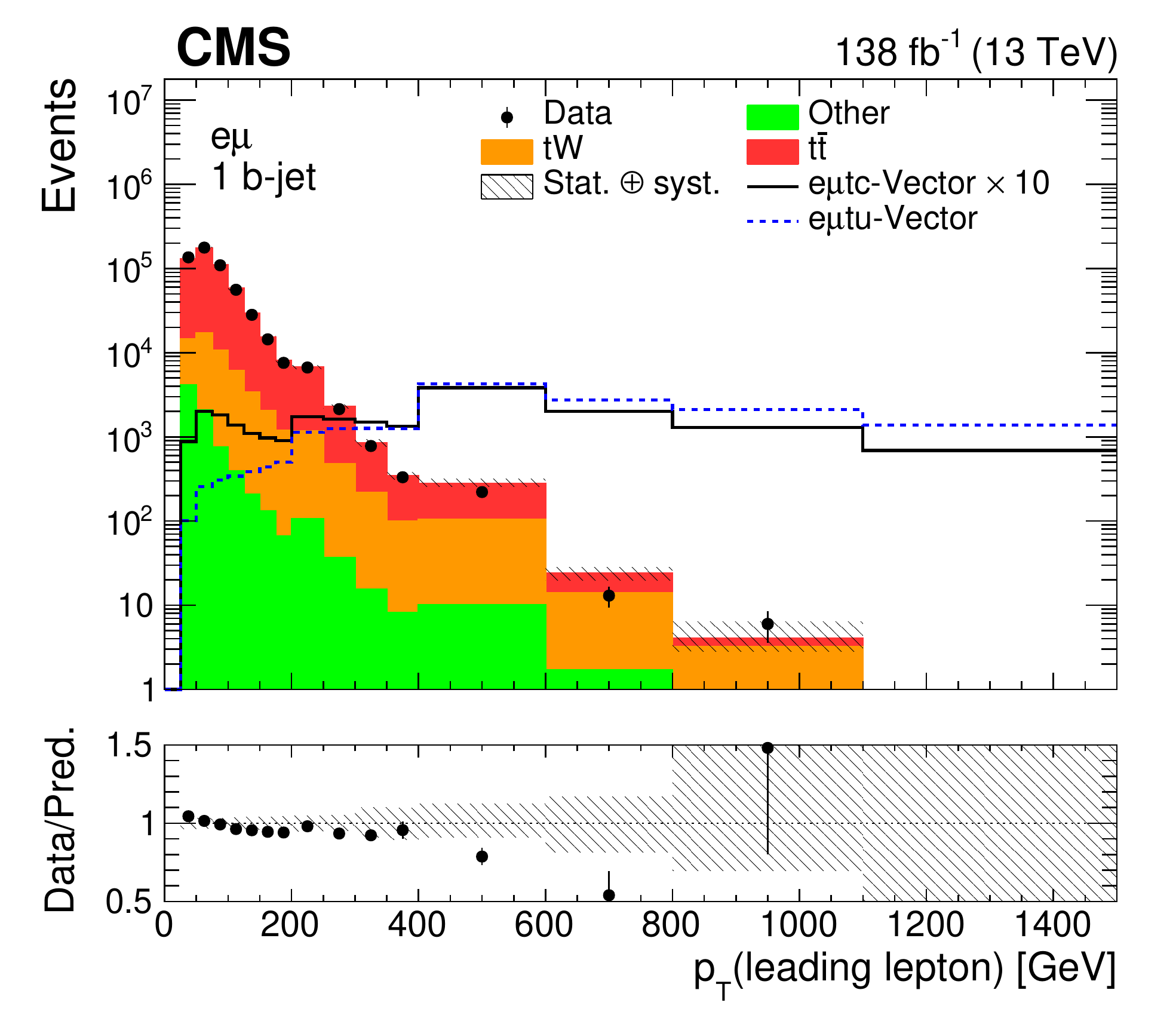}
      \includegraphics[width=0.44\textwidth]{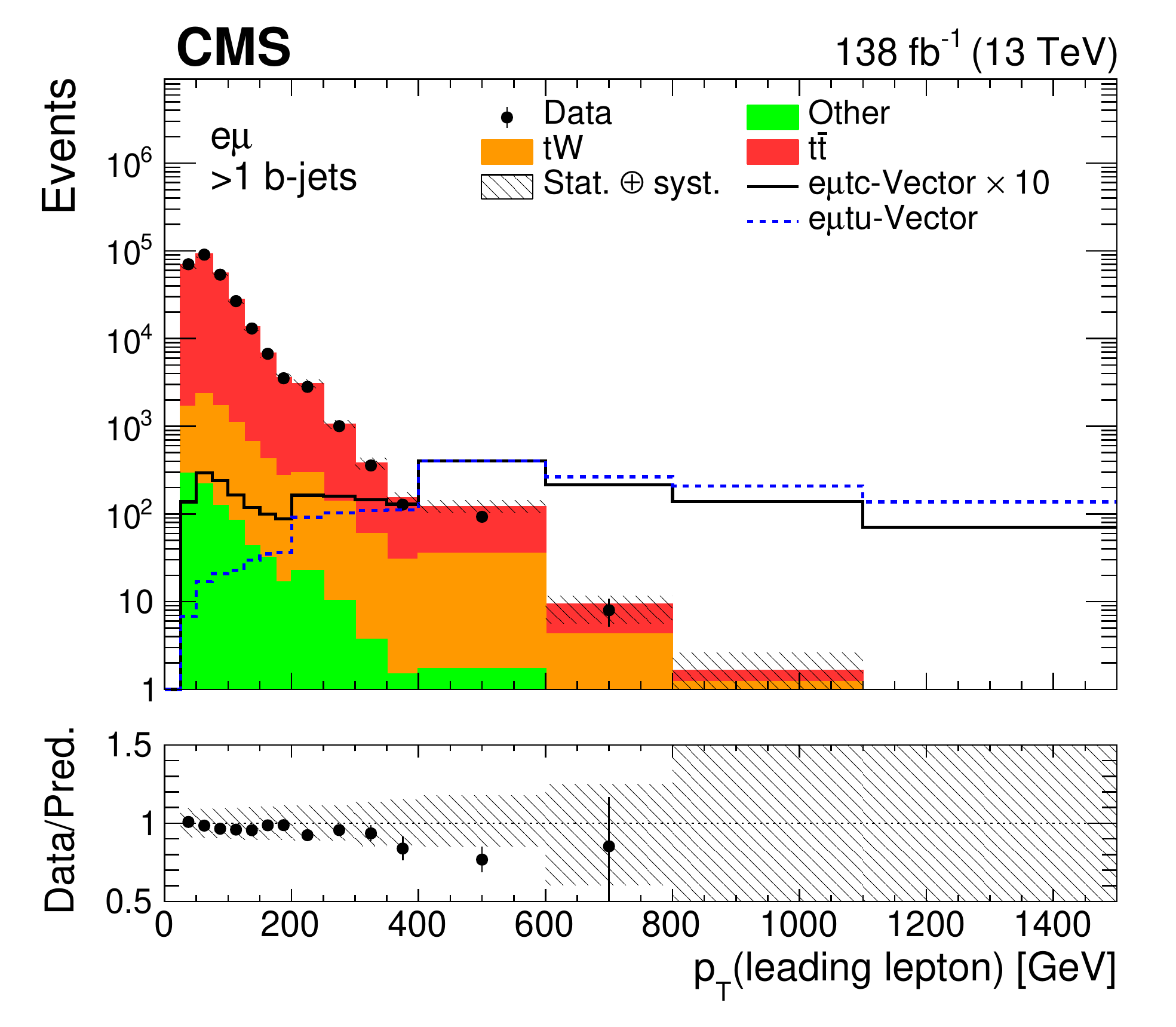}
      \includegraphics[width=0.44\textwidth]{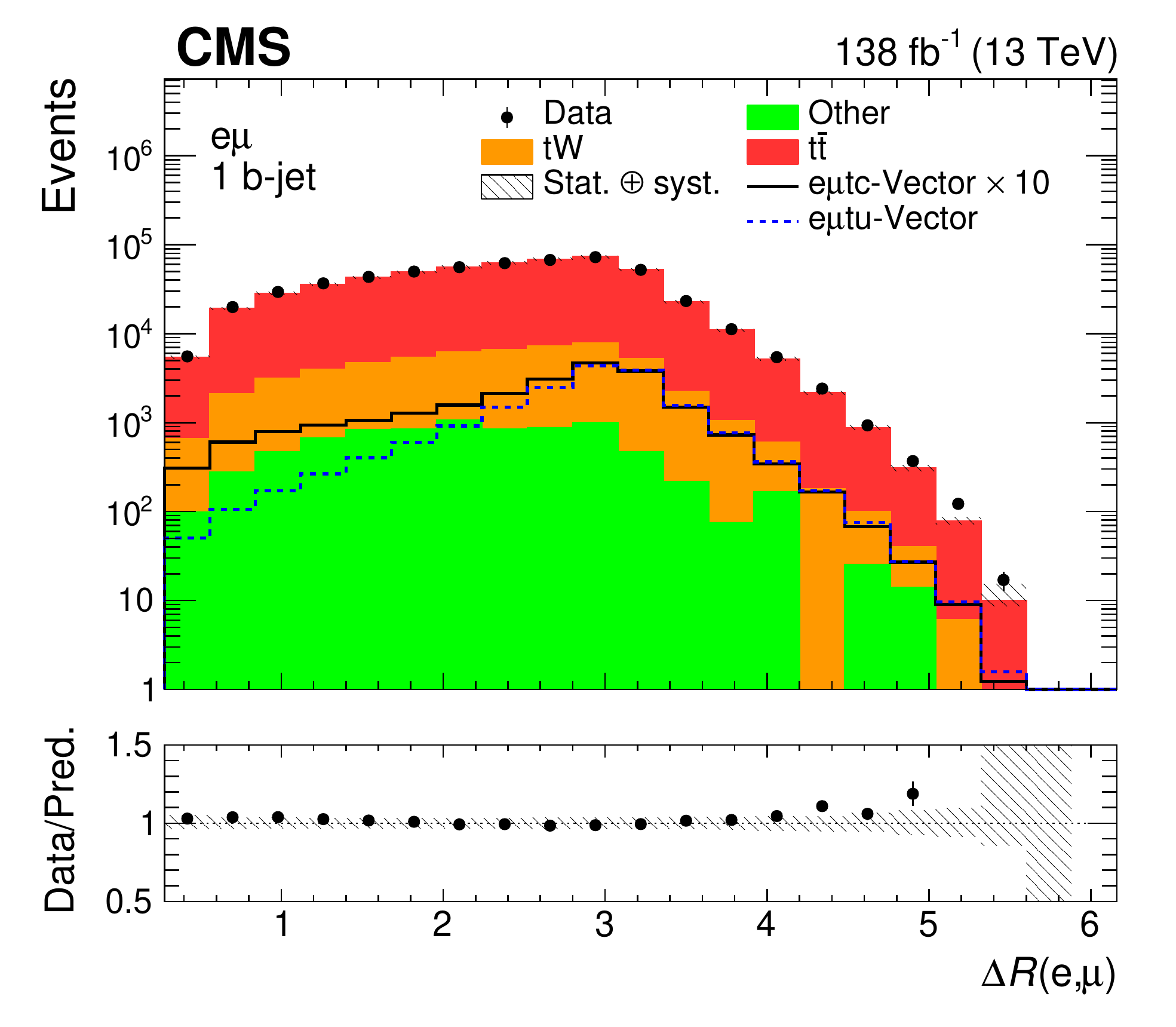}
      \includegraphics[width=0.44\textwidth]{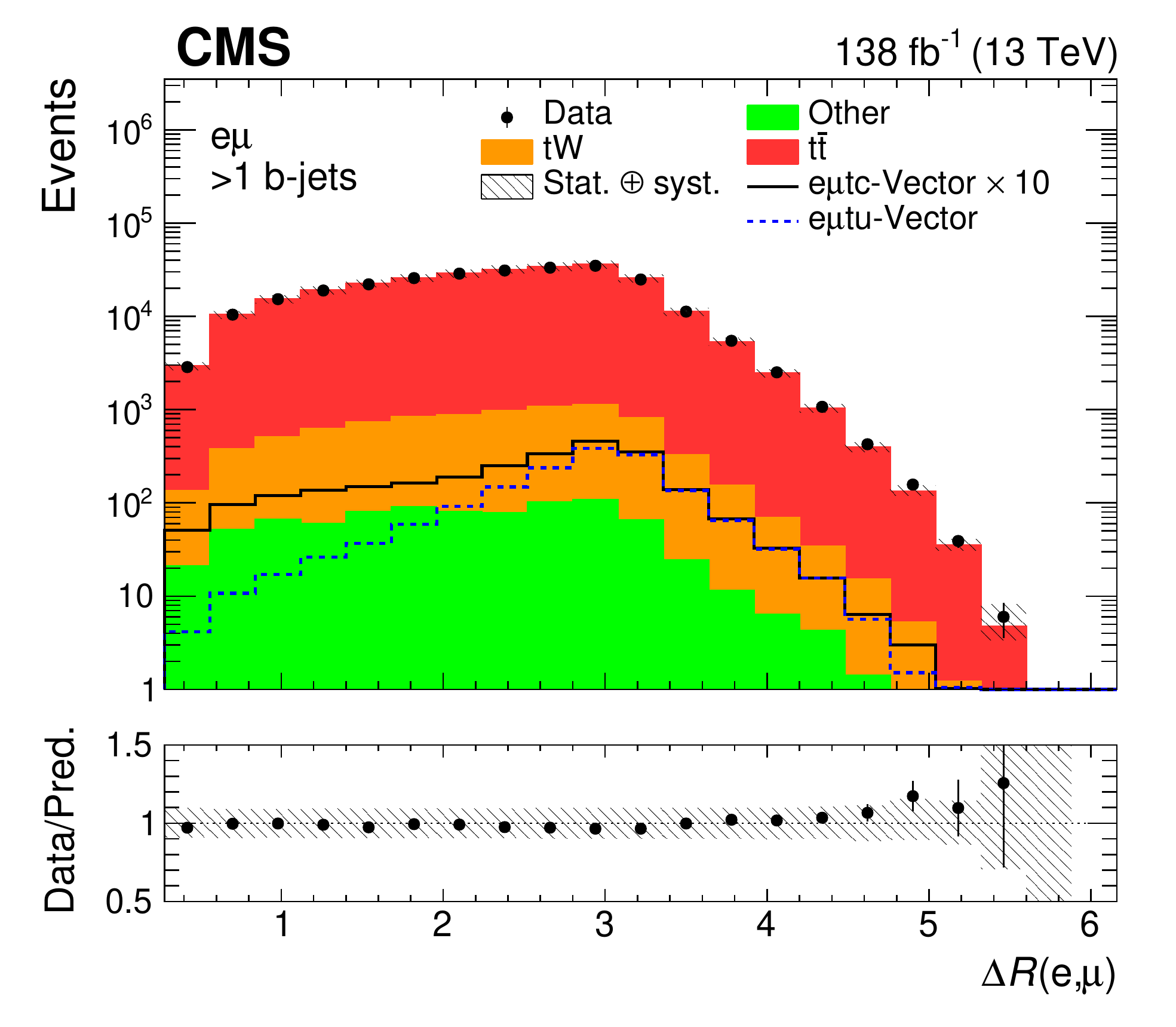}
      \includegraphics[width=0.44\textwidth]{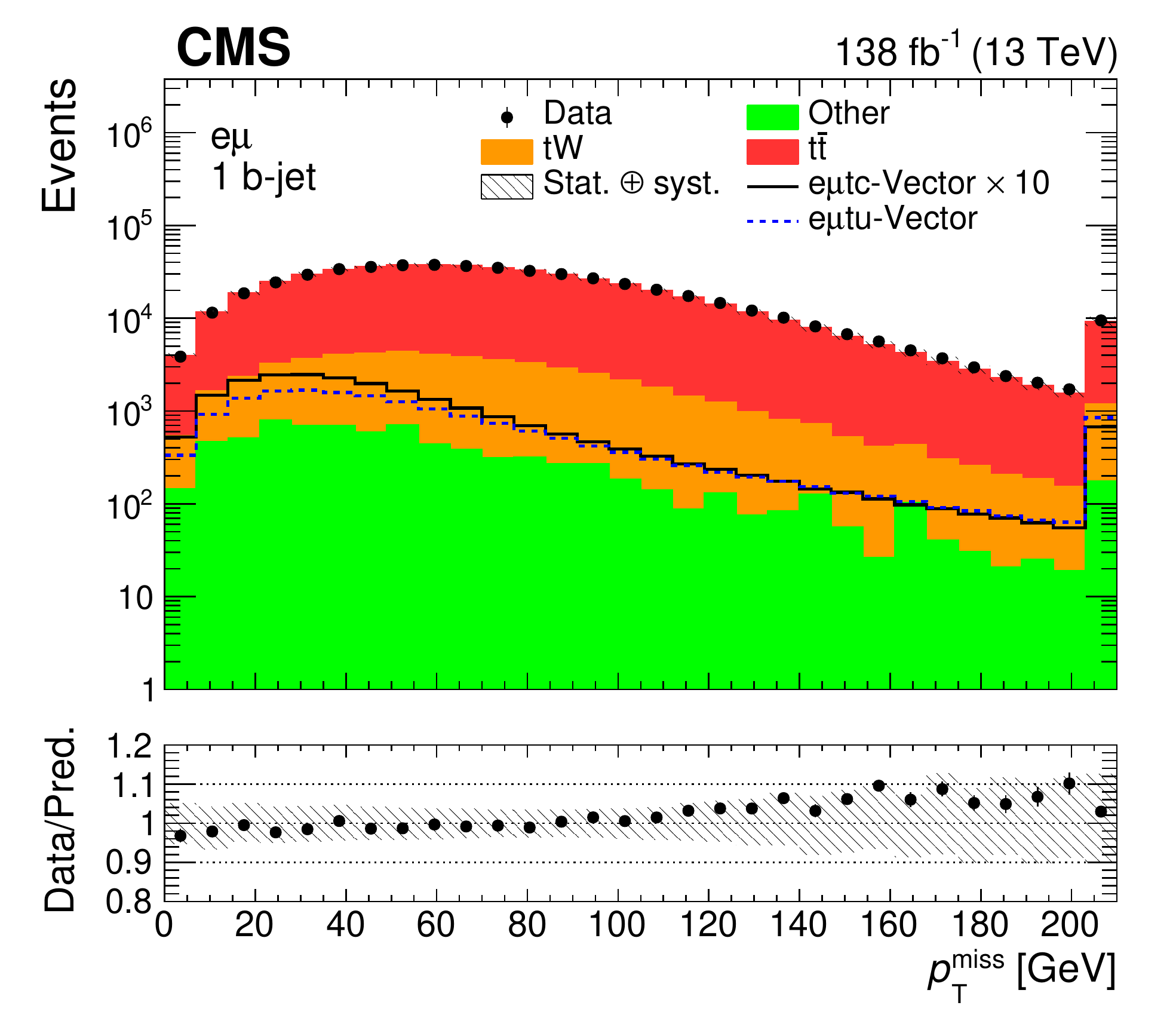}
      \includegraphics[width=0.44\textwidth]{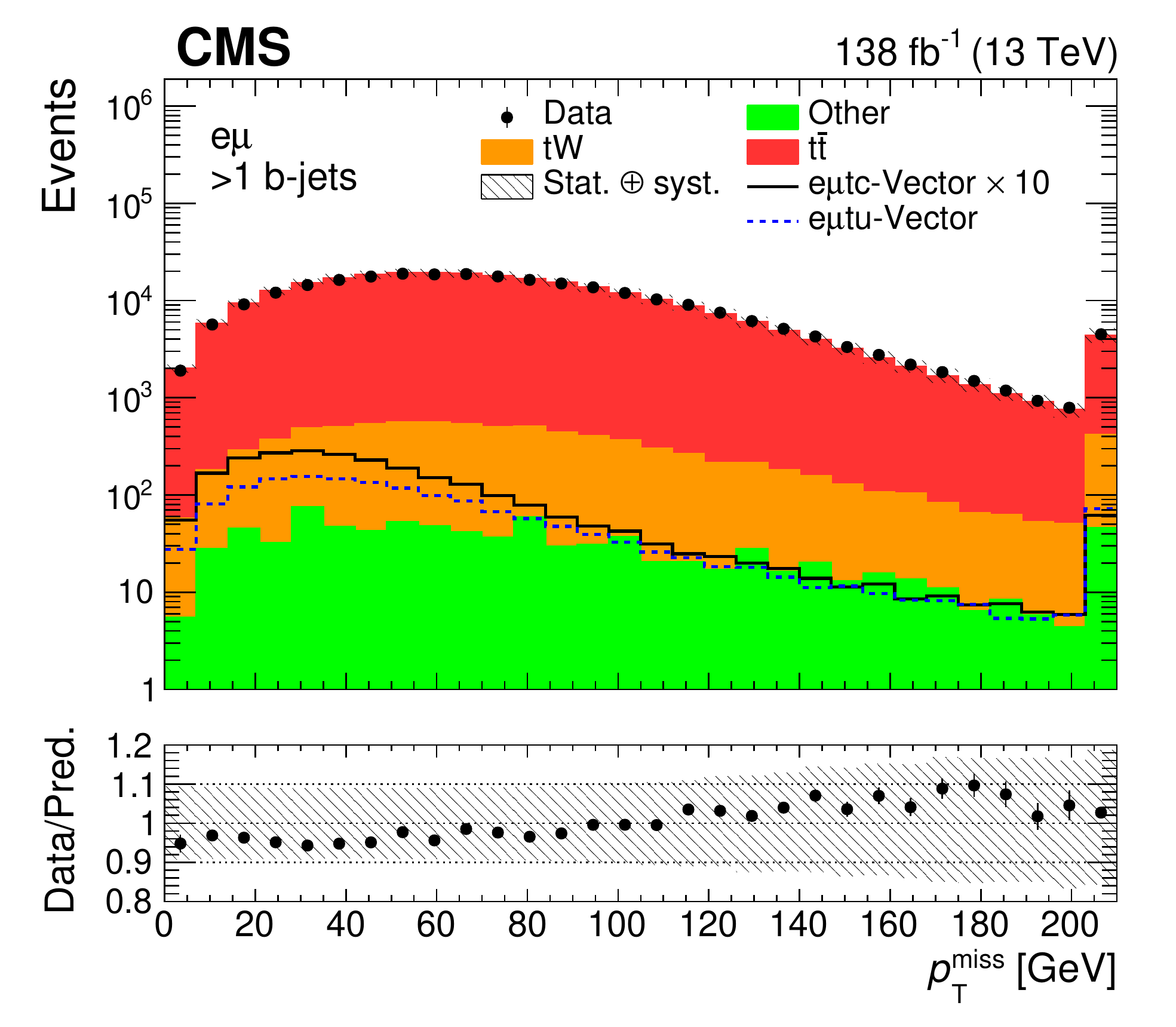}

  \caption{ The distributions of the leading lepton \pt (upper row), $\Delta R(\Pe, \PGm)$ (middle row), and \ptmiss (lower row) are shown for data (points) and simulation (histograms). 
Events in the signal region (one \cPqb tagged jet) and \ttbar control region (more than one \cPqb tagged jets) are shown in the left and right column, respectively.
The hatched bands indicate the total uncertainty (statistical and systematic taken in quadrature) for the SM background predictions (cf. Section~\ref{sec:unc}).
Overflow events are added to the last bin.
Examples of the predicted signal contribution for the vector type CLFV interactions via $\Pe\PGm \PQt \PQu$ and $\Pe\PGm \PQt \PQc$ vertices are shown, assuming $\cxlbd$.
The signal production- and decay-mode contributions are summed.
The $\Pe\PGm \PQt \PQc$ signal cross section is scaled up by a factor of 10 for improved visualization.
}
  \label{fig:BDTinput1}
  \end{center}
\end{figure}

\begin{figure}[hp!]
  \begin{center}
      \includegraphics[width=0.44\textwidth]{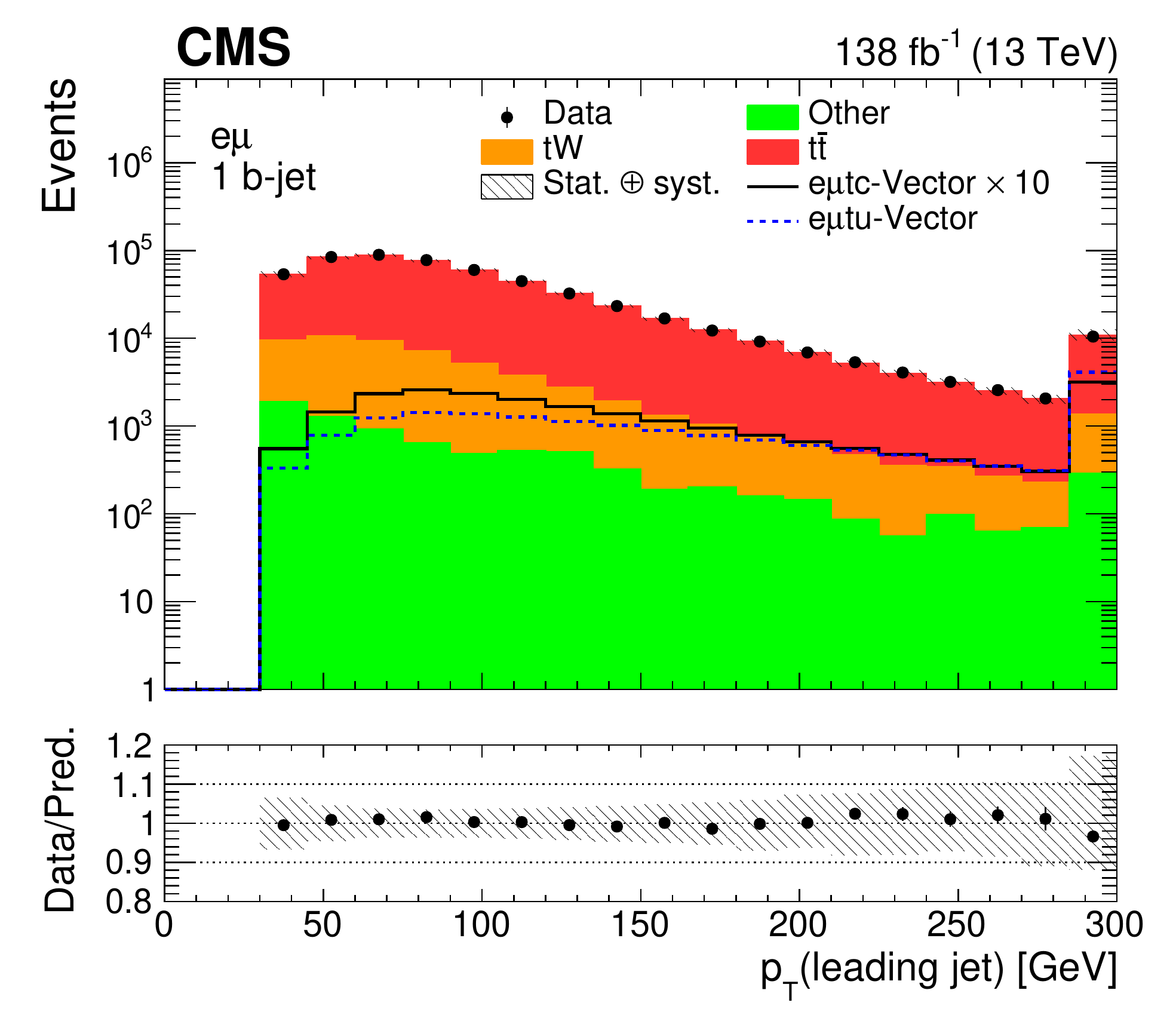}
      \includegraphics[width=0.44\textwidth]{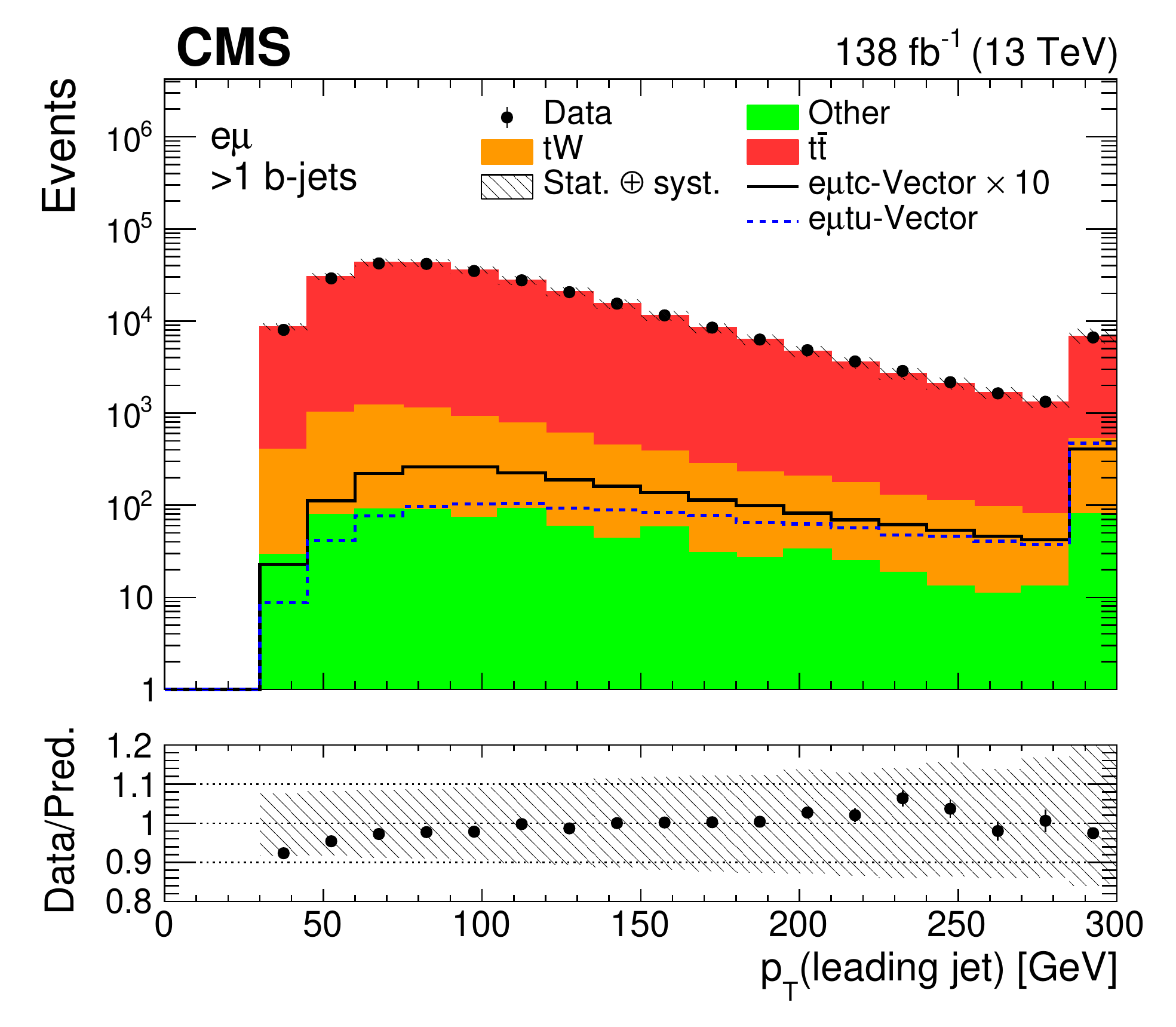}
      \includegraphics[width=0.44\textwidth]{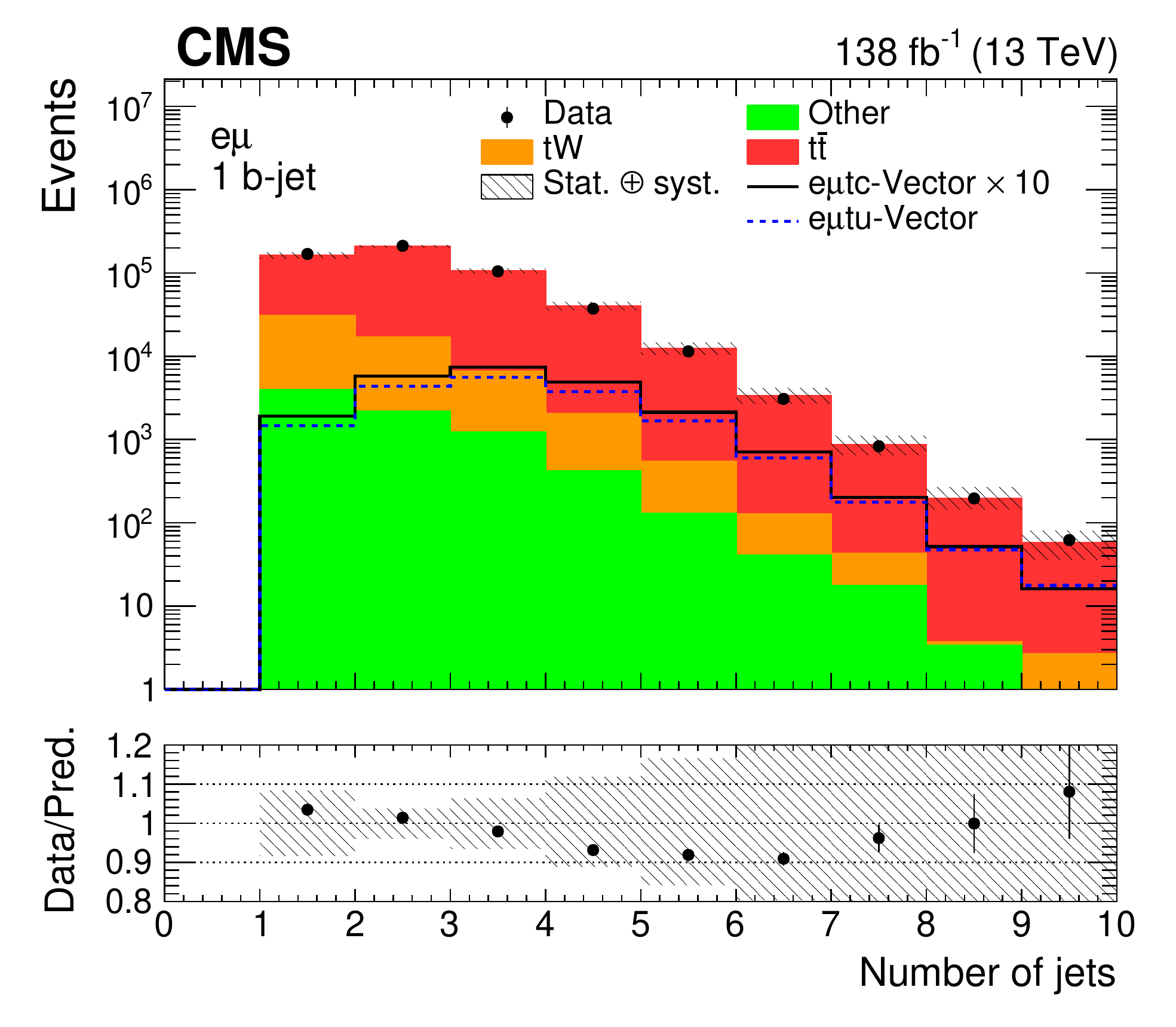}
      \includegraphics[width=0.44\textwidth]{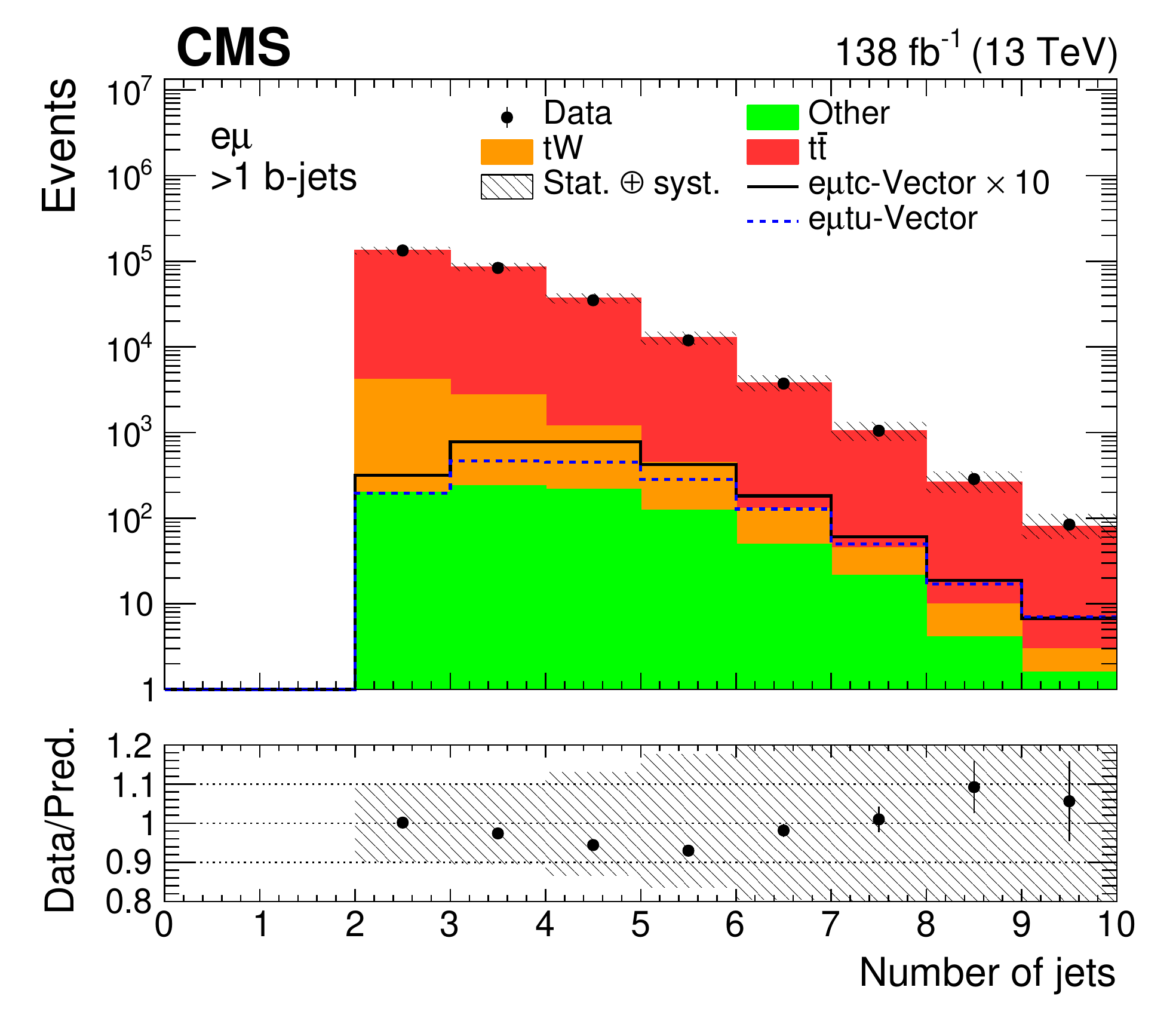}
  \caption{ The distributions of the leading jet \pt (upper row) and the number of jets (lower row) are shown for data (points) and simulation (histograms). 
Events in the signal region (one \cPqb tagged jet) and \ttbar control region (more than one \cPqb tagged jets) are shown in the left and right column, respectively.
The hatched bands indicate the total uncertainty (statistical and systematic taken in quadrature) for the SM background predictions (cf. Section~\ref{sec:unc}).
Overflow events are added to the last bin.
Examples of the predicted signal contribution for the vector type CLFV interactions via $\Pe\PGm \PQt \PQu$ and $\Pe\PGm \PQt \PQc$ vertices are shown, assuming $\cxlbd$.
The signal production- and decay-mode contributions are summed.
The $\Pe\PGm \PQt \PQc$ signal cross section is scaled up by a factor of 10 for improved visualization.
}
  \label{fig:BDTinput2}
  \end{center}
\end{figure}

The BDT uses 5 variables: the \pt of the leading lepton ($\pt^{\ell_1}$ where $\ell$ refers to \Pe or \Pgm),
the \pt of the leading jet, the distance between the electron and muon [$\Delta R(\Pe, \Pgm) = \sqrt{\smash[b]{( \eta^{\Pe} - \eta^{\Pgm})^2 +  (\phi^{\Pe} - \phi^{\Pgm})^2}}$],
\ptmiss, and the number of jets. 
Figures~\ref{fig:BDTinput1} and \ref{fig:BDTinput2} provide distributions of the BDT input variables in data and simulations for signal and \ttbar control regions.
A good description of the data is observed for the background model.
The leading lepton distribution is  somewhat softer in data, although it is within the estimated systematic uncertainties after \pt reweighting of simulated SM \ttbar events to the most precise cross section available (cf. Section~\ref{sec:samples}) \cite{Sirunyan:2018ucr}.

The CLFV single top quark production and top quark decay events, weighted according to their cross sections, are compared against the SM \ttbar events in the BDT training. 
The BDT is trained and tested on independent samples with no evidence of overtraining or bias.
As shown in Table~\ref{cutflow}, the CLFV single top quark production channel has higher yields than the CLFV top quark decay channel in all signal samples.
In addition, events from the CLFV single top quark production channel result in higher \pt on average for the final-state particles when compared to the CLFV decay channel.
Therefore, events from the CLFV single top quark production channel play a leading role in the BDT discrimination.
The vector, scalar, and  tensor CLFV samples show similar distributions in the selected BDT input variables.
A single BDT is therefore trained using all signal samples in the region with one \PQb tagged jet, and is used to probe all of the CLFV Wilson coefficients.
To control the background uncertainties in the fit, the trained BDT in the signal region is used in the \ttbar control region.

\clearpage

\section{Systematic uncertainties}
\label{sec:unc}

Various sources of systematic uncertainty affect  the final signal and background yields and distributions.
The systematic uncertainties are categorized into two classes: experimental uncertainties arising from modeling of the detector response, and theoretical uncertainties arising from the modeling of the signal and background processes in the MC simulation. 
The uncertainties that do not depend on run conditions, such as theoretical uncertainties, are treated as correlated across the different data-taking periods. 
The systematic uncertainties for the three data-taking years are treated as correlated unless noted otherwise.

Lepton reconstruction, identification, and isolation efficiencies are determined using $\PZ\to \ell \ell$ events, and scale factors (SF) are applied to all MC simulations to correct any discrepancies between data and simulation~\cite{Khachatryan:2015hwa,Chatrchyan:2012xi}. 
The SFs depend on lepton \pt and $\eta$.
The uncertainties in lepton momentum scale and resolution are computed by changing the simulated \pt by their uncertainties, and then repeating the analysis \cite{Sirunyan:2019yvv,Sirunyan:2020ycc}. 
The trigger efficiency in data is measured through a \ptmiss requirement since the efficiency of the \ptmiss trigger is independent of the dilepton trigger.  
The SFs are applied to account for the differences in trigger efficiencies between data and simulation as functions of leading and sub-leading lepton \pt.
The trigger uncertainty is estimated by changing the trigger scale factors by their uncertainties originating from sample size, event topology, and lepton SFs \cite{Sirunyan:2018ucr}.
The trigger uncertainty is considered uncorrelated among different years.

The uncertainty arising from the jet energy scale is calculated from 27 sources, 
where each source is either fully correlated or uncorrelated among the years.
Each source of uncertainty depends on jet \pt and $\eta$ \cite{Khachatryan:2016kdb}. The uncertainty in jet energy resolution is considered uncorrelated among different years. 
The quantity \ptmiss is recalculated whenever the jet momenta are rescaled to estimate its uncertainty. 
An additional uncertainty in the calculation of \ptmiss is estimated through changes in  the energies of reconstructed particles that are not clustered into jets by their respective resolutions, and then recalculating the \ptmiss.
The uncertainties associated with \PQb tagging are determined by changing the related SFs by one standard deviation~\cite{Sirunyan:2017ezt}.
These uncertainties depend on the \pt of each jet and amount to approximately 1--5\% per jet.

The integrated luminosities of the 2016, 2017, and 2018 data-taking periods are individually known with uncertainties in the 1.2--2.5\% range~\cite{CMS-LUM-17-003,LUM-17-004,LUM-18-002}, while the total Run-2 (2016--2018) integrated luminosity has an uncertainty of 1.6\%, the improvement in precision reflecting the reduction
of errors when combining some uncorrelated systematic effects.
The uncertainty associated with modeling pileup is estimated by changing the \pp total inelastic cross section by $\pm 4.6\%$ \cite{Sirunyan:2018nqx}.

The impact of theoretical assumptions in modeling signal or SM \ttbar background  is determined by repeating the analysis and replacing the nominal events by dedicated simulation samples with altered parameters or
by changing the reference simulation using the source-related weights.
The uncertainty arising from missing higher-order QCD terms in the simulation of the signal and SM \ttbar processes at matrix-element (ME) level is assessed by changing the renormalization and factorization scales up and down by factors of two relative to the nominal values.  Unphysical cases, where one scale fluctuates up while the other fluctuates down,
are not considered. 
The uncertainty related to the choice of PDF is evaluated using replicas of the NNPDF3.0 and NNPDF3.1 parameters~\cite{Ball:2014uwa,Ball:2017nwa,Kalogeropoulos:2018cke}.
Uncertainties in initial- and final-state QCD radiation (ISR and FSR) are evaluated by changing the renormalization scale for QCD emissions in ISR and FSR up and down by a factor of 2.
The three mentioned sources of modeling uncertainties are considered for both signal and SM \ttbar processes. 
In addition, uncertainties originating from the scheme used to match the ME-level calculation  to the parton-shower (PS) simulation, the modeling of the underlying event defined in \PYTHIA tunes (UE tune), and the models of color reconnection for the SM \ttbar process according to what is described in Ref. \cite{Sirunyan:2018ucr} are included.
The \ttbar and signal modeling uncertainties refer to the impact on the acceptance only.
For uncertainties related to SM \ttbar, \tw, and other background contributions, we use normalization uncertainties of 5, 10, and 30\% \cite{Sirunyan:2018ucr,Sirunyan:2018lcp}.  

The systematic uncertainties in signal and SM \ttbar selection efficiencies are summarized in Table~\ref{sysTab}. 
The largest uncertainty is from the \PQb tagging SF since we have used SFs that are measured in inclusive multijet samples instead of dilepton \ttbar events to reduce a potential bias.
Although only a representative signal sample is shown in Table~\ref{sysTab}, all signal samples have similar uncertainties. 
Except the uncertainties in total integrated luminosities and background normalizations, all other uncertainties affect both the background rate and the shape of the BDT distributions.

\begin{table}[htbp!]
\topcaption{Summary of representative systematic uncertainties in selection efficiency for the SM \ttbar process and for single top quark production and decays via vector $\Pe\PGm \PQt\PQu$ CLFV interactions in the signal plus \ttbar control regions.}
\centering
\begin{tabular}{lrrr}
\hline
Source                                     & \ttbar (\%)    & \multicolumn{2}{c}{CLFV signal}    \\  
                                           &                & decay (\%)            & production (\%)     \\ 
\hline
Trigger                                    & 1.2     & 1.2            & 2.9         \\  
Electron identification and isolation      & 1.6     & 1.6            & 3.9         \\  
Muon identification and isolation          & 0.6     & 0.6            & 0.7         \\  
Electron energy scale and resolution       & $<$0.1 & $<$0.1        & $<$0.1     \\  
Muon momentum scale and resolution         & $<$0.1 & $<$0.1        & $<$0.1     \\  
Jet energy scale and resolution            & 2.5     & 2.1            & 1.2         \\
\PQb tagging                              & 3.1     & 3.9            & 4.5         \\
Pileup                                     & 0.3     & 0.3            & 0.2         \\
ME scale                                   & 0.9     & 0.8            & 0.7         \\
ISR/FSR scale                              & 1.5     & 2.9            & 1.9         \\
PDF                                        & 0.8     & 0.8            & 0.9         \\
UE tune                                    & 0.4     & \NA            & \NA         \\
ME/PS matching                             & $<$0.1 & \NA            & \NA         \\
Color reconnection                         & 1.0     & \NA            & \NA         \\  
MC statistical                             & $<$0.1 & $<$0.1         & $<$0.1    \\
\hline
\end{tabular}
\label{sysTab}
\end{table}

\section{Results}
\label{sec:results}

\begin{figure}[hp!]
  \begin{center}
      \includegraphics[width=0.48\textwidth]{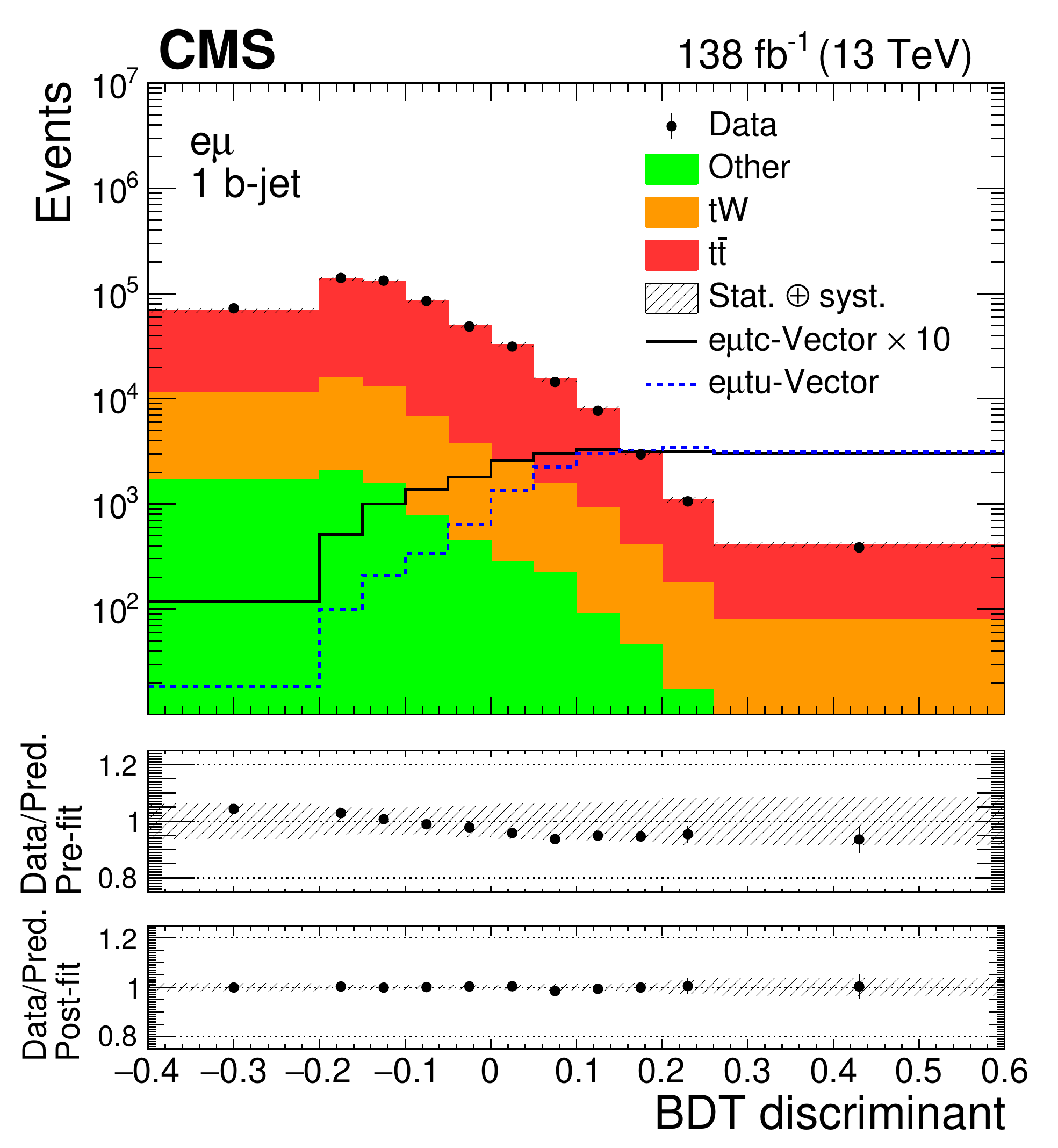}
      \includegraphics[width=0.48\textwidth]{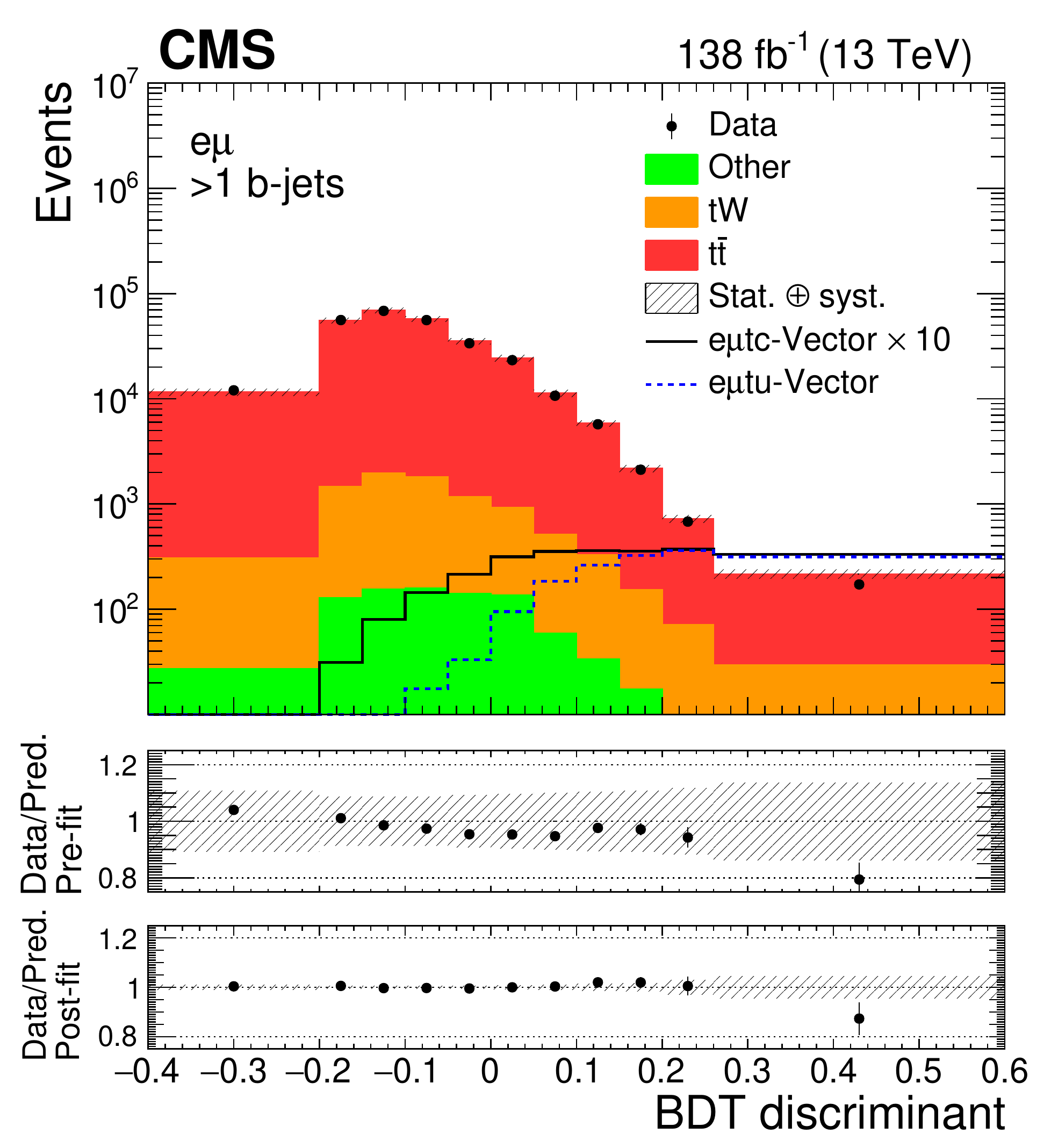}
  \caption{The BDT output distributions for data (points) and backgrounds (histograms)  with the ratio of data to the
total background yield, before (middle panel) and after (lower panel) the fit.
Events in the signal region (one \cPqb tagged jet) and \ttbar control region (more than one \cPqb tagged jets) are shown in the left and right column, respectively.
The hatched bands indicate the total uncertainty (statistical and systematic taken in quadrature) for the SM background predictions (cf. Section~\ref{sec:unc}).
Examples of the predicted signal contribution for the vector type CLFV interactions via $\Pe\PGm \PQt \PQu$ and $\Pe\PGm \PQt \PQc$ vertices are shown, assuming $\cxlbd$.
The signal production- and decay-mode contributions are summed.
The $\Pe\PGm \PQt \PQc$ signal cross section is scaled up by a factor of 10 for improved visualization.
}
  \label{fig:BDT}
  \end{center}
\end{figure}

The final BDT discriminant distributions for the three data-taking years and two data regions (signal region and \ttbar control region) are jointly used to test for the presence of signal events.
A binned likelihood function $\mathcal{L}(\mu,\theta)$  constructed as a product of Poisson probability terms over all bins is used for the statistical analysis 
where $\mu$ is the signal-strength parameter and $\theta$ is a set of nuisance parameters. 
The parameter of interest, $\mu$, changes the cross sections of both signal channels, top quark CLFV production and decay, by exactly the same scale.
The cross sections of both signal channels depend quadratically on the CLFV Wilson coefficients. Since our signal samples are normalized to the cross sections at $\cxlbd$, $\sqrt{\mu}$ and $\text{C}_x/\Lambda^2$ are equivalent parameters.
All the systematic uncertainties defined in Section~\ref{sec:unc} are treated as nuisance parameters $\theta$, assuming a log normal prior for normalization parameters, and Gaussian priors for BDT shape uncertainties.
The uncertainties due to the limited number of simulated events used for  signal and background expectations are taken into account using ``the Barlow--Beeston lite'' method \cite{BARLOW1993219}.
The data are found to be consistent with expectations of the SM in the absence of signal.
The observed distributions of the BDT discriminant, together with the SM background expectations, before and after a fit to  signal plus background hypothesis are shown in Fig.~\ref{fig:BDT}. 

\begin{figure}[htp!]
  \begin{center}
      \includegraphics[width=0.48\textwidth]{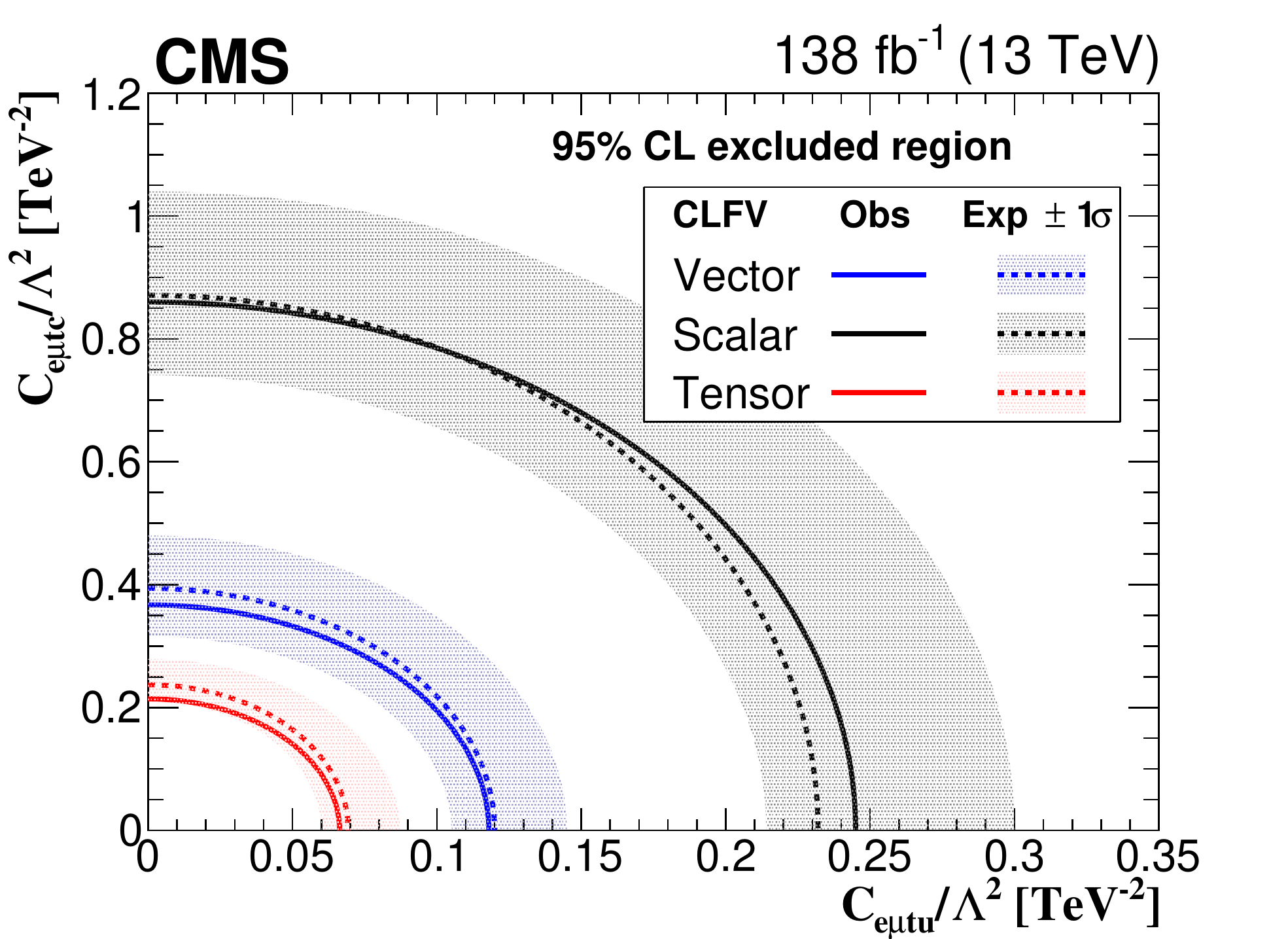}
      \includegraphics[width=0.48\textwidth]{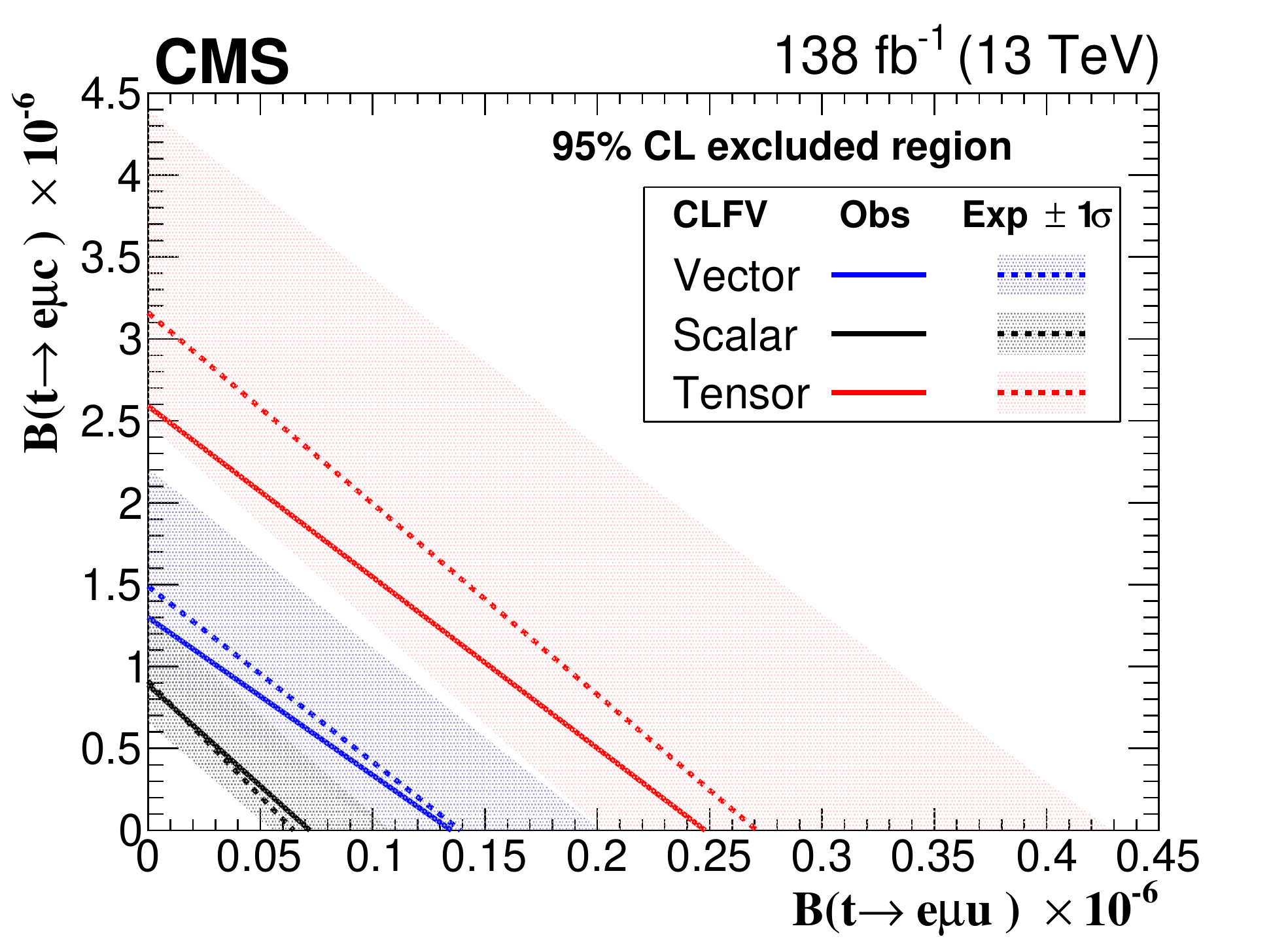}
  \caption{The observed 95\% CL exclusion limits on the $\Pe\PGm \PQt \PQc$; of the $\Pe\PGm \PQt \PQu$ Wilson coefficient (left) and $\mathcal{B}(\PQt \to \Pe\PGm \PQc)$ as a function of $\mathcal{B}(\PQt \to \Pe\PGm \PQu)$ (right) for the vector-, scalar-, and tensor-like CLFV interactions. 
The hatched bands indicate the regions containing 68\% of the distribution of limits expected under the background-only hypothesis. 
}
  \label{fig:limit}
  \end{center}
\end{figure}

Upper limits on the production cross section for signal are set at 95\% confidence level (\CL) using the modified frequentist \CLs method~\cite{JUNK1999435, Read_2002}, with a likelihood ratio as a test statistic.
The limit setting procedure is performed for a given individual Wilson coefficient ($\cvector$, $\cscalar$, or $\ctensor$) while the other Wilson coefficients are set to zero.
Consequently, upper limits on the Wilson coefficients are translated to limits on the related top quark CLFV branching fractions \cite{Kile:2008rp}.
Limits obtained for vector-, scalar- and tensor-like interactions are summarized in Table~\ref{Rfull}.
The measured one-dimensional exclusion limits are also interpreted for the scenario of
the non-vanishing $\Pe\PGm \cPqt\cPqu$ and $\Pe\PGm \cPqt\cPqc$ CLFV couplings via a linear interpolation. The results for two-dimensional
limits on CLFV Wilson coefficients and branching fractions are displayed in Fig. \ref{fig:limit}.
The sources of systematic uncertainty with the largest impact on the estimated signal contribution depend on the CLFV interaction type.
The three main sources of uncertainty that are common among the CLFV interaction types  are uncertainties in SM \ttbar FSR, electron SFs, and the normalization  of the SM \ttbar  process.
The other backgrounds have a negligible influence on the limits.

The limit obtained on the tensor CLFV Wilson coefficient is more stringent than those on scalar and vector coefficients because of its larger relative production cross section, as presented in  Table~\ref{cutflow}.
Tabulated results are provided in HEPDATA \cite{hepdata}.
When translated into limits on the branching fractions to CLFV final states, the relative contributions of the tensor and scalar operators to the decay translate into more stringent limits on the scalar operators \cite{Kile:2008rp}. 

\begin{table}[htbp!]
\centering
\topcaption{Expected and observed 95\% \CL upper limits on the CLFV Wilson coefficients and top quark CLFV branching fractions. 
}
\label{Rfull}
\begin{tabular}{cccccc}
\hline \rule{0pt}{2.6ex}
 Vertex & Int.  & \multicolumn{2}{c}{$\text{C}_{\Pe\PGm \PQt \PQq}/\Lambda^2$ [\TeVns$^{-2}$]}  & \multicolumn{2}{c}{$\mathcal{B} (10^{-6})$} \\
                       & type   &Exp         &Obs  &Exp            &Obs    \\ \hline  
 \multirow{3}{*}{$\Pe\PGm \PQt\PQu$}   & Vector        & 0.12 & 0.12 & 0.14 & 0.13     \\
                       & Scalar        & 0.23 & 0.24 & 0.06 & 0.07     \\
                       & Tensor        & 0.07 & 0.06 & 0.27 & 0.25     \\[\cmsTabSkip]
 \multirow{3}{*}{$\Pe\PGm \PQt\PQc$}                      & Vector        & 0.39 & 0.37 & 1.49 & 1.31     \\
                       & Scalar        & 0.87 & 0.86 & 0.91 & 0.89     \\
                       & Tensor        & 0.24 & 0.21 & 3.16 & 2.59     \\ \hline
\end{tabular}
\end{table}

\section{Summary}
\label{sec:summary}

A search is reported for charged-lepton flavor violation in top quark production and decay. 
The analysis is based on \pp collisions collected by the CMS detector at the LHC at a center-of-mass energy of 13 \TeV, corresponding to an integrated luminosity of 138\fbinv.
Events are selected if they contain an oppositely charged electron-muon pair and at least one \PQb tagged jet.
An effective field theory approach is used for parametrizing top quark lepton flavor violating interactions.
The production and decay modes of the top quark through these effective interactions are included in this analysis. 

A boosted decision tree is used to distinguish signal from background. 
No significant excess is observed over the expectations from the standard model.
Upper limits are set on the strength of the individual vector-, scalar-, and tensor-like  four-fermion effective operators.
These are converted to limits on the branching fractions of the top quark $\mathcal{B}(\PQt \to \Pe\PGm \PQq)$, \PQq = \PQu (\PQc) quark, ${<}0.13\times 10^{-6}$ ($1.31\times 10^{-6}$), $0.07\times 10^{-6}$ ($0.89\times 10^{-6}$), and $0.25\times 10^{-6}$ ($2.59\times 10^{-6}$) for vector, scalar, and tensor CLFV interactions, respectively.
The resulting limits are the most restrictive bounds to date.

\begin{acknowledgments}

  We congratulate our colleagues in the CERN accelerator departments for the excellent performance of the LHC and thank the technical and administrative staffs at CERN and at other CMS institutes for their contributions to the success of the CMS effort. In addition, we gratefully acknowledge the computing centers and personnel of the Worldwide LHC Computing Grid and other centers for delivering so effectively the computing infrastructure essential to our analyses. Finally, we acknowledge the enduring support for the construction and operation of the LHC, the CMS detector, and the supporting computing infrastructure provided by the following funding agencies: BMBWF and FWF (Austria); FNRS and FWO (Belgium); CNPq, CAPES, FAPERJ, FAPERGS, and FAPESP (Brazil); MES and BNSF (Bulgaria); CERN; CAS, MoST, and NSFC (China); MINCIENCIAS (Colombia); MSES and CSF (Croatia); RIF (Cyprus); SENESCYT (Ecuador); MoER, ERC PUT and ERDF (Estonia); Academy of Finland, MEC, and HIP (Finland); CEA and CNRS/IN2P3 (France); BMBF, DFG, and HGF (Germany); GSRI (Greece); NKFIA (Hungary); DAE and DST (India); IPM (Iran); SFI (Ireland); INFN (Italy); MSIP and NRF (Republic of Korea); MES (Latvia); LAS (Lithuania); MOE and UM (Malaysia); BUAP, CINVESTAV, CONACYT, LNS, SEP, and UASLP-FAI (Mexico); MOS (Montenegro); MBIE (New Zealand); PAEC (Pakistan); MSHE and NSC (Poland); FCT (Portugal); JINR (Dubna); MON, RosAtom, RAS, RFBR, and NRC KI (Russia); MESTD (Serbia); MCIN/AEI and PCTI (Spain); MOSTR (Sri Lanka); Swiss Funding Agencies (Switzerland); MST (Taipei); ThEPCenter, IPST, STAR, and NSTDA (Thailand); TUBITAK and TAEK (Turkey); NASU (Ukraine); STFC (United Kingdom); DOE and NSF (USA).

\hyphenation{Rachada-pisek} Individuals have received support from the Marie-Curie program and the European Research Council and Horizon 2020 Grant, contract Nos.\ 675440, 724704, 752730, 758316, 765710, 824093, 884104, and COST Action CA16108 (European Union); the Leventis Foundation; the Alfred P.\ Sloan Foundation; the Alexander von Humboldt Foundation; the Belgian Federal Science Policy Office; the Fonds pour la Formation \`a la Recherche dans l'Industrie et dans l'Agriculture (FRIA-Belgium); the Agentschap voor Innovatie door Wetenschap en Technologie (IWT-Belgium); the F.R.S.-FNRS and FWO (Belgium) under the ``Excellence of Science -- EOS" -- be.h project n.\ 30820817; the Beijing Municipal Science \& Technology Commission, No. Z191100007219010; the Ministry of Education, Youth and Sports (MEYS) of the Czech Republic; the Deutsche Forschungsgemeinschaft (DFG), under Germany's Excellence Strategy -- EXC 2121 ``Quantum Universe" -- 390833306, and under project number 400140256 - GRK2497; the Lend\"ulet (``Momentum") Program and the J\'anos Bolyai Research Scholarship of the Hungarian Academy of Sciences, the New National Excellence Program \'UNKP, the NKFIA research grants 123842, 123959, 124845, 124850, 125105, 128713, 128786, and 129058 (Hungary); the Council of Science and Industrial Research, India; the Latvian Council of Science; the Ministry of Science and Higher Education and the National Science Center, contracts Opus 2014/15/B/ST2/03998 and 2015/19/B/ST2/02861 (Poland); the Funda\c{c}\~ao para a Ci\^encia e a Tecnologia, grant CEECIND/01334/2018 (Portugal); the National Priorities Research Program by Qatar National Research Fund; the Ministry of Science and Higher Education, projects no. 0723-2020-0041 and no. FSWW-2020-0008, and the Russian Foundation for Basic Research, project No.19-42-703014 (Russia); MCIN/AEI/10.13039/501100011033, ERDF ``a way of making Europe", and the Programa Estatal de Fomento de la Investigaci{\'o}n Cient{\'i}fica y T{\'e}cnica de Excelencia Mar\'{\i}a de Maeztu, grant MDM-2017-0765 and Programa Severo Ochoa del Principado de Asturias (Spain); the Stavros Niarchos Foundation (Greece); the Rachadapisek Sompot Fund for Postdoctoral Fellowship, Chulalongkorn University and the Chulalongkorn Academic into Its 2nd Century Project Advancement Project (Thailand); the Kavli Foundation; the Nvidia Corporation; the SuperMicro Corporation; the Welch Foundation, contract C-1845; and the Weston Havens Foundation (USA).

\end{acknowledgments}

\bibliography{auto_generated}

\cleardoublepage \appendix\section{The CMS Collaboration \label{app:collab}}\begin{sloppypar}\hyphenpenalty=5000\widowpenalty=500\clubpenalty=5000\input{TOP-19-006-authorlist.tex}\end{sloppypar}
\end{document}

%% file: TOP-19-006-authorlist.tex
\cmsinstitute{Yerevan~Physics~Institute, Yerevan, Armenia}
A.~Tumasyan
\cmsinstitute{Institut~f\"{u}r~Hochenergiephysik, Vienna, Austria}
W.~Adam\cmsorcid{0000-0001-9099-4341}, J.W.~Andrejkovic, T.~Bergauer\cmsorcid{0000-0002-5786-0293}, S.~Chatterjee\cmsorcid{0000-0003-2660-0349}, M.~Dragicevic\cmsorcid{0000-0003-1967-6783}, A.~Escalante~Del~Valle\cmsorcid{0000-0002-9702-6359}, R.~Fr\"{u}hwirth\cmsAuthorMark{1}, M.~Jeitler\cmsAuthorMark{1}\cmsorcid{0000-0002-5141-9560}, N.~Krammer, L.~Lechner\cmsorcid{0000-0002-3065-1141}, D.~Liko, I.~Mikulec, P.~Paulitsch, F.M.~Pitters, J.~Schieck\cmsAuthorMark{1}\cmsorcid{0000-0002-1058-8093}, R.~Sch\"{o}fbeck\cmsorcid{0000-0002-2332-8784}, D.~Schwarz, S.~Templ\cmsorcid{0000-0003-3137-5692}, W.~Waltenberger\cmsorcid{0000-0002-6215-7228}, C.-E.~Wulz\cmsAuthorMark{1}\cmsorcid{0000-0001-9226-5812}
% ...
\vskip\cmsinstskip
% ...
1:~Also at TU Wien, Wien, Austria\\
% ...
